\documentclass{emulateapj}

\usepackage{graphicx}
\usepackage{amsmath}

\usepackage{graphicx}
\usepackage{amsmath}

\newcommand{\appropto}{\mathrel{\vcenter{
  \offinterlineskip\halign{\hfil$##$\cr
    \propto\cr\noalign{\kern2pt}\sim\cr\noalign{\kern-2pt}}}}}

\shortauthors{Kalomeni et al.~2016}
\shorttitle{Evolution of CVs}

\begin{document}

\bibliographystyle{apj}

\title{Evolution of Cataclysmic Variables and Related Binaries Containing a White-Dwarf}

\author{
B.~Kalomeni\altaffilmark{1,2},
L.~Nelson\altaffilmark{3},
S.~Rappaport\altaffilmark{1},
M.~Molnar\altaffilmark{1},
J.~Quintin\altaffilmark{4,5},
K.~Yakut\altaffilmark{2}
}

\altaffiltext{1}{Department of Physics, and Kavli Institute for Astrophysics and Space Research, Massachusetts Institute of Technology, Cambridge, MA 02139, USA; kalomeni@mit.edu, sar@mit.edu, momchil.molnar@gmail.com, }

\altaffiltext{2}{Department of Astronomy and Space Sciences, Ege University, 35100, \.Izmir, Turkey; belinda.kalomeni@ege.edu.tr, kadri.yakut@ege.edu.tr}

\altaffiltext{3}{Department of Physics, Bishop's University, 2600 College St., Sherbrooke, Quebec, QC J1M 1Z7, Canada; lnelson@ubishops.ca}

\altaffiltext{4}{Department of Physics, McGill University, Montr\'eal, QC H3A 2T8, Canada;  jquintin@physics.mcgill.ca}

\altaffiltext{5}{Vanier Canada Graduate Scholar}

\slugcomment{Submitted to the {\it Astrophysical Journal}, 2016 September 29}

\begin{abstract}

We present a binary evolution study of cataclysmic variables (CVs) and related systems with white dwarf accretors, including for example, AM CVn systems, classical novae, supersoft X-ray sources, and systems with giant donor stars.  Our approach intentionally avoids the complications associated with population synthesis algorithms thereby allowing us to present the first truly comprehensive exploration of all of the subsequent binary evolution pathways that ZACVs might follow (assuming fully non-conservative, Roche-lobe overflow onto an accreting WD) using the sophisticated binary stellar evolution code MESA.  The grid consists of 56,000 initial models, including 14 white dwarf accretor masses, 43 donor-star masses ($0.1-4.7$ $M_{\odot}$), and 100 orbital periods. We explore evolution tracks in the orbital period and donor-mass ($P_{\rm orb}-M_{\rm don}$) plane in terms of evolution dwell times, masses of the white dwarf accretor, accretion rate, and chemical composition of the center and surface of the donor star.  We report on the differences among the standard CV tracks, those with giant donor stars, and ultrashort period systems.  We show where in parameter space one can expect to find supersoft X-ray sources, present a diagnostic to distinguish among different evolutionary paths to forming AM CVn binaries, quantify how the minimum orbital period in CVs depends on the chemical composition of the donor star, and update the $P_{\rm orb}(M_{\rm wd})$ relation for binaries containing white dwarfs whose progenitors lost their envelopes via stable Roche-lobe overflow.  Finally, we indicate where in the $P_{\rm orb}-M_{\rm don}$ the accretion disks will tend to be stable against the thermal-viscous instability, and where gravitational radiation signatures may be found with LISA.

\end{abstract}


\section{Introduction}

Cataclysmic variables (CVs) are widely characterized as interacting binaries consisting of an accreting white dwarf (WD) in close proximity to a mass-losing star (see, e.g., Patterson 1984 and Warner 1995 for comprehensive reviews). According to the conventional model for the formation of CVs in the disk of our galaxy, the progenitors are wide primordial binaries for which the more massive star evolves to become an RGB/AGB star\footnote{Other evolutionary pathways involving stable mass transfer are possible.}.  As its radius expands, and assuming that mass transfer is dynamically unstable, the companion star spirals in towards the core of the giant thereby stripping it of its envelope and leaving the companion in an increasingly tighter orbit with a WD star (i.e., the core of the primary). This short-lived dynamical phase is known as common envelope (`CE') evolution (see, e.g., Paczy{\'n}ski 1976; Taam et al.~1978; Webbink~1984; Zorotovic et al.~2010; Passy et al.~2011). Assuming that a merger does not occur, it is possible for the companion (donor) to subsequently fill its Roche lobe and initiate mass transfer (the incipient zero-age cataclysmic variable [`ZACV']). The post common envelope binary (`PCEB') can only reach this stage in its evolution if the timescales for angular momentum dissipation (gravitational radiation/magnetic braking) and/or nuclear evolution of the donor are sufficiently short (see, e.g., Pylyser \& Savonije 1988).

According to the standard model for CV evolution, the mass of the donor star continuously decreases (unless magnetic braking is abruptly interrupted) together with a reduction in the orbital period ($P_{\rm orb}$). When the mass of the donor has been reduced to $\lesssim 0.1 \, M_\odot$, the thermal timescale of the donor becomes sufficiently long and its interior partially electron degenerate so that further reduction in mass causes the orbital period to increase (see, e.g., Paczy{\'n}ski \& Sienkiewicz 1981; Rappaport, Joss \& Webbink 1982 [RJW]; Rappaport, Verbunt \& Joss 1983 [RVJ]; Nelson et al.~1985; Ritter 1985; Hameury et al.~1988). This behavior leads to a `minimum orbital period' of approximately 80 minutes which has been observed in the present-day population of CVs (PDCVs) (see, e.g., G{\"a}nsicke et al.~2009).

Although there are several distinct features such as the orbital period minimum and the orbital period gap between 2 to 3 hours (see, e.g., RVJ) that are the hallmarks of the CV population, they are incredibly rich in terms of the diversity of the physical phenomena that they exhibit. Moreover, they play an important role in chemical enrichment (e.g., classical novae; see Livio \& Truran 1994) and may even be important contributors to the frequency of Type Ia supernovae (Maoz 2011; Chen et al.~2014).  Many of the groups that comprise the population of CVs are classified either in terms of the behavior of the accretion disks surrounding the WD accretors or in terms of the chemical composition and rate of mass transfer onto the accreting WDs (including the intrinsic properties of the WDs themselves). Some of the groupings include: dwarf novae (e.g., the U Gem, Z Cam, and SU Ursa Majoris subclasses), polars (including DQ Her and AM Her stars), supersoft X-ray sources (SXSs), novalike variables, recurrent novae (RNe), classical novae (CNe), SW Sextantis stars, and the ultracompact AM CVn binaries (see, e.g., Patterson 1984; Iben 1991; Warner 1995; Greiner 1996; Ritter \& Kolb 2003; Eggleton 2006; Knigge et al. 2014).

There has been considerable progress in the development of a unified picture to explain the relationships among all these groups but the picture is far from complete. Recent observational surveys such as SDSS have provided much needed data and helped build our understanding (see, e.g., G{\"a}nsicke et al.~2009;  Szkody et al.~2011), but some very important questions remain unanswered.   For example, can any significant fraction of the observed short orbital period AM CVn binaries actually be derived from the CV population (see Podsiadlowski et al.~2003; Bildsten et al.~2007; Townsley \& Bildsten~2007; van der Sluys et al.~2009; Nelemans et al. 2010). Could nova-like variables and SW Sex stars be related to short timescale phenomena such as extreme thermal timescale mass transfer or perhaps even bordering on a dynamical timescale? Did a significant fraction of PDCVs actually descend from much higher mass donors than we observe today? Other unsatisfactorily answered questions include the relationship among CNe, RNe, and SXSs.

These issues are intimately connected to the problem of reproducing the observed mass distribution of the accreting white dwarfs in CVs. It is generally agreed that WDs in CVs show a tendency to be more massive ($> 0.7M_\odot$, see, e.g., Littlefair et al.~2008, and Zorotovic et al.~2011) than the expected theoretical value that  includes both canonical CO WDs with masses of $\sim$$0.6 \,M_\odot$ and even lower-mass helium WDs.
If observational selection effects are not severe, then the question arises as to how the observed WDs have significantly more mass. It is possible that WDs grow in mass and clearly the rate of mass-transfer is a key factor in determining the growth (Nomoto 1982, 1988; van den Heuvel et al.~1992; Zorotovic et al.~2011). A rigorous calculation of this growth has proven problematic because of a lack of agreement on how the growth-rate depends on the accretion-rate. A theoretical simulation carried out by Wijnen et al.~(2015) concludes that it is not possible to replicate the observed WD mass distribution in CVs purely as a result of mass accretion. Instead, Schreiber et al.~(2015), and later Nelemans et al.~(2016), invoke (empirical) consequential angular momentum losses (CAML) to explain the mass distribution and to reconcile the observationally inferred spatial density of CVs with the theoretical predictions. The CAML mechanism is thought to cause many of the CVs with low-mass WDs to become dynamically unstable and thus to merge into a single object. This substantially reduces the number of CVs containing low-mass WDs thereby increasing the average WD mass in CVs and, moreover, it lowers the spatial density of CVs in the galactic disk yielding better agreement with observational inferences.

In order to compute the frequency of the formation of the various types of CV systems described above, population synthesis techniques have been used to obtain quantitative estimates. Some of the important population synthesis studies that have helped contribute to our understanding of CVs include those by de Kool (1992), Kolb (1993), Politano (1996), Howell et al.~(1997), Nelemans et al. (2001a), Howell et al.~(2001), Podsiadlowski et al.~(2003),  Nelson et al.~(2004a), Politano (2004), Chen et al.~(2014), Goliasch \& Nelson~(2015), and Wijnen et al.~(2015). However, the relative complexity of the physical processes and the wide range of timescales associated with population synthesis simulations still pose considerable difficulties even with currently available computing power. The large number of dimensions of parameter space has often required that simplifying assumptions be implemented in order to make the computations tractable. Except for the pioneering study by Podsiadlowski et al.~(2003) and the recent works of Chen et al.~(2014) and Goliasch \& Nelson~(2015), previous population synthesis studies have not fully included the effects of the internal chemical evolution of the donor star on the PDCV population. Moreover, no study has fully accounted for CV systems that are born above the `bifurcation limit' (see Pylyser \& Savonije 1988; see also Sects.~\ref{sec:illustrative} and \ref{sec:results}). These are the systems for which the nuclear timescale of the donors is so short that they can form helium or CO cores before sufficient mass is stripped away allowing them to eventually evolve to much longer orbital periods. The evolutionary endpoints of these systems are usually detached double degenerates, but while evolving to that state they should resemble high-mass CVs.

While population synthesis techniques are extremely powerful, there are still a number of unresolved issues that require further investigation (see the discussion above). These difficulties are likely due to two reasons: (i) observational selection effects; and, (ii) inadequate physical descriptions of all the processes that are relevant to the properties of present-day CVs starting from the birth of the primordial binaries to the long-term evolution of CVs themselves. One of the limitations of these studies is that the probability rules that govern the frequencies of various evolutionary channels are usually quite rigid. For example, the probability that primordial main-sequence binaries form with particular masses and/or mass ratios is prescribed. Even though a range of probability parameters may be analyzed in the investigation, it may not be feasible to examine all possible evolutionary channels because they have in essence been implicitly excluded. This would not be a problem if probability rules were known with some high degree of precision.

In principle, if we had a statistically large and unbiased sample of CVs, pre-CVs and PCEBs, we could compare all possible evolutionary pathways with the observations and `solve' the inverse problem.  This would allow us to conclude which physical mechanisms are dominant at every stage of the system's evolutionary history.  Thus based on the properties measured for any present-day CV, these constraints might possibly allow us to infer the initial conditions for the corresponding ZACV.  We could also determine the subsequent evolution of the system if the properties of the PCEBs have been precisely measured (Ritter~1986; Schreiber \& G{\"a}nsicke 2003; Davis et al.~2010).  As an example, consider the system TYC 6760-497-1 which appears to be the first pre-SXS ever discovered (Parsons et al.~2015). The comprehensive grid of evolutionary models presented in this work would allow us to identify which evolutionary tracks most closely match its currently observed properties.

Our approach is to avoid the complications associated with population synthesis algorithms and to present the first truly comprehensive exploration of all of the subsequent binary evolution pathways that ZACVs might follow (assuming fully non-conservative, Roche-lobe overflow onto an accreting WD) using a sophisticated binary stellar evolution code MESA (Paxton et al. 2011).  Thus our methodology is similar to that described by Lin et al.~(2011) who started with a grid of initial conditions (i.e., the masses of the neutron-star accretor and the donor in addition to $P_{\rm orb}$) and subsequently followed the evolution of low-mass X-ray binaries and the formation of binary radio pulsars.  The binaries in our grid are assumed to have formed from primordial systems (as described above) but it is also possible to imagine that they could have formed in some other way (e.g., tidal capture in dense stellar environments that led to rapid circularization before the onset of mass transfer). Specifically, we have run about 56,000 binary models covering a uniform grid of 14 WD accretors with masses, $M_{\rm wd}$, between 0.1 and 1.4 $M_\odot$, 40 main-sequence donor masses, $M_{\rm don}$, between 0.3 and 4.7 $M_\odot$, and 100 initial orbital periods logarithmically distributed between 10 and 485 hours. The evolutions are then followed over a duration of $10^{10}$ years. Since we assume no priors in terms of the probability constraints on the initial binary component masses and orbital periods, we are in essence assuming that all of the initial conditions of our starting grid are equally likely. This allows us to construct unbiased analyses of the observationally relevant planes such as the $P_{\rm orb}-M_{\rm don}$, the $\dot M-M_{\rm don}$, and the $\log L- \log T_{\rm eff}$ planes. Our tracks evolve through regions of all known types of accreting WD systems{\footnote {Magnetic CVs are believed to follow similar evolutionary tracks to non-magnetic CVs, but it should be noted that they may continue to lose mass as they evolve through the period gap.}} (except symbiotic binaries and pathological cases such as may occur due to dynamical interactions in clusters), but also reveal some new unexplored possible types of WD accreting systems which could be searched for observationally. We also study the likely end products for each type of evolution.

As an example of the utility of our approach, we determine what subset of initial conditions leads to the formation of ultracompact AM CVn type systems  (Podsiadlowski et al.~2003) and what the range of our initial conditions is that leads to the formation of double WD systems (Kilic et al. 2011; Brown et al.~2011; Kilic et al.~2014; Maxted et al.~2014). Moreover, we also deduce the type of degenerate companion that is created once the system becomes detached (e.g., whether it is a HeWD or COWD). We also attempt to address the question as to whether a substantial fraction of PDCVs are derived from high mass ($\gtrsim 1.5 \,M_\odot$) primordial donors. We make robust predictions as to how the ratio of observable isotopes such as N/O would have to change in order for us to be able to conclude that a particular PDCV descended from a high-mass primordial donor. Observations of the abundance ratios for C/O and N/O inferred from the chemical properties of the donor star and the accretion disk in CVs have been carried out by Harrison et al.~(2004), Howell et al.~(2010), and Hamilton et al.~(2011). They found evidence for carbon deficiencies in CVs that would seem to imply that some CVs for which they measured the abundances could have been derived from much more massive systems. Details concerning the evolution of the surface abundances of the CNO isotopes are analyzed in this work.

The paper is organized as follows. In Sect.~\ref{sec:MESA} we discuss the stellar evolution code used for the binary evolution calculations, and spell out some of the assumptions used. The grid of initial binary parameters is described in Sect.~\ref{sec:grid}. In order for the reader to better understand the large number of evolution tracks we present, in Sect.~\ref{sec:illustrative} we first present and describe a smaller illustrative set of the binary evolutions. Here we explain why the tracks split into three basic branches: (i) conventional CVs, (ii) systems with donor stars on the giant branch, and (iii) ultracompact systems with orbital periods down to $\sim$5 minutes. The results of our evolution study are presented in Sect.~\ref{sec:results}. Color-coded images are presented in the orbital period-donor mass plane where the color indicates a Ôthird parameterÕ of the system (e.g., evolution time, donor mass, mass transfer rate, composition, etc.). An HR diagram for all the evolution tracks is presented in Appendix A.  We discuss the minimum in the orbital periods of CVs, ultracompact systems, and systems undergoing thermal timescale mass transfer in Sect.~\ref{sec:Pmin}.  The final states of the binaries at the end of their evolution, including the relation between orbital period and relic core mass of the donor star, are discussed in Sect.~\ref{sec:final}. The orbital period distribution corresponding to our evolution tracks is presented in Sect.~\ref{sec:dist}.  Regions of accretion disk stability are discussed in Sect.~\ref{sec:disk_stability}, where we present the ratio of the accretion rate to the critical rate for stability in the $P_{\rm orb}-M_{\rm don}$ plane.  We calculate gravitational wave strain amplitudes ($h$) for all of our evolutionary tracks and also display them in the $P_{\rm orb}-M_{\rm don}$ plane in Sect.~\ref{sec:gravity}. In Sect.~\ref{sec:discuss} we show where the observed accreting white dwarf systems fall on our evolution diagrams and discuss the surface hydrogen abundances of the donors.  Finally, in Sect.~\ref{sec:summary} we summarize our results and draw some final conclusions.

\section{Binary Stellar Evolution Code}
\label{sec:MESA}

In this work we use the MESA stellar evolution code (Paxton et al.~2011; 2013; 2015) to model the evolution of the donor star as it loses mass to a white-dwarf accretor. The specific version of MESA was 3851.  The evolution of the binary is handled by the MESA test suite {\tt rlo}, which basically allows for mass loss from the donor star and systemic loss of angular momentum due to magnetic braking (see, e.g., Rappaport et al.~1983), gravitational radiation, and mass loss from the system.  This current study is analogous in a number of ways to that carried out for neutron star accretors by Lin et al.~(2011), and further details of the input and prescriptions used for the binary evolution calculations can be found therein.

\begin{figure}
\centering
\includegraphics[angle=-0, width=0.98\columnwidth]{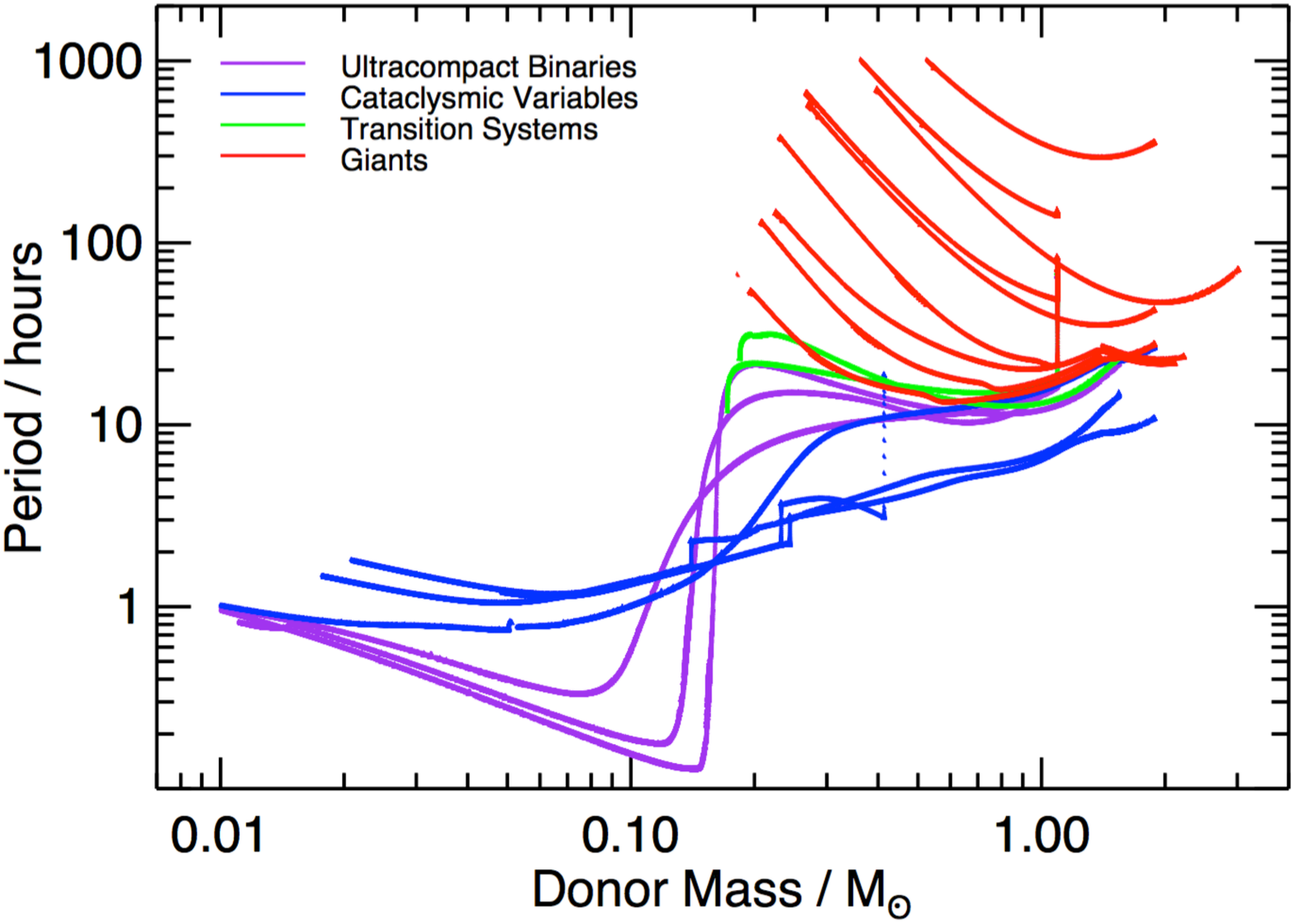}
\caption{Set of illustrative evolution tracks in the orbital period--donor-mass plane.  Blue, purple, red, and green curves are for classical CVs, ultracompact systems, systems with giant donor stars, and transitional systems between the ultracompact and giant systems.  A variety of white dwarf accretor masses are used, including 1.4, 1.0, 0,7, and 0.5 $M_\odot$.}
 \label{fig:scheme}
\end{figure}

The evolution time steps for the donor star itself are controlled by MESA (Paxton et al.~2011).  In terms of the binary orbital evolution, the time steps are set so that the fractional changes per time step in the donor-star mass and the orbital period are limited to $\Delta M_{\rm don}/M_{\rm don} < 0.003$ and $\Delta P_{\rm don}/P_{\rm don} < 0.003$, respectively. Magnetic braking (Verbunt \& Zwaan 1981; RVJ) is turned on whenever the donor star has a radiative core and when its mass is $M_{\rm don} < 1.4\,M_\odot$, i.e., when the star does not have a significant radiative envelope.  The binary evolution is stopped at $10^{10}$ years, when the mass of the donor drops below 0.01 $M_\odot$, or when $P_{\rm orb} > 40$ days, whichever comes first.

The mass retention fraction of the white dwarf is taken to be zero, and the lost matter carries away the specific angular momentum of the white dwarf accretor.  The justification for this is that the accreting white dwarf is assumed to expel most, if not all, of its accreted mass either in nova explosions at lower rates of accretion (i.e., $\lesssim$ few\,$\times 10^{-8} \, M_\odot $ yr$^{-1}$;  Gallagher \& Starrfield 1978; Prialnik et al.~1982; Bildsten et al.~2007), or in an intense radiatively driven stellar wind at the higher rates of accretion (i.e., $\gtrsim 10^{-6} \, M_\odot$ yr$^{-1}$; Kato \& Hachisu 1988; Hachisu \& Kato 2001).  Even in the narrow accretion-rate regime where steady-state nuclear burning can take place (i.e., $5 \times 10^{-8} - 5 \times 10^{-7} \, M_\odot$ yr$^{-1}$;  Paczy{\'n}ski  \& Rudak 1980; Iben 1982; Nomoto 1982, 1988; Iben \& Tutukov 1989), we assume that eventually the accumulated He shell becomes unstable and explodes.  Our assumption of zero mass retention by the WD, which would preclude the formation of Type Ia supernovae for a small fraction of the systems, does not significantly affect the overall evolution of these systems.  The reason, in a nutshell, is that the orbital period of a semi-detached binary depends only on the mass and radius of the Roche-lobe filling star (the `donor') and not on the mass of the companion accreting star (so long as the donor is less massive than the accretor). Finally, for reference, we note that the Eddington limited accretion rate for a $\sim$$1 \, M_\odot$ white dwarf is about $10^{-5} \,M_\odot$ yr$^{-1}$.

Once the envelope of a donor star that is ascending the giant branch is lost, the MESA evolution of its remnant core continues.  Lower-mass degenerate He cores mostly cool to their final degenerate state, while for cores with $M \gtrsim 0.31 \, M_\odot$, the He core can continue to burn to C and O, thereby forming HeCO hybrid white dwarfs.

\begin{figure}
  \centering
\includegraphics[angle=0, width=1.00\columnwidth]{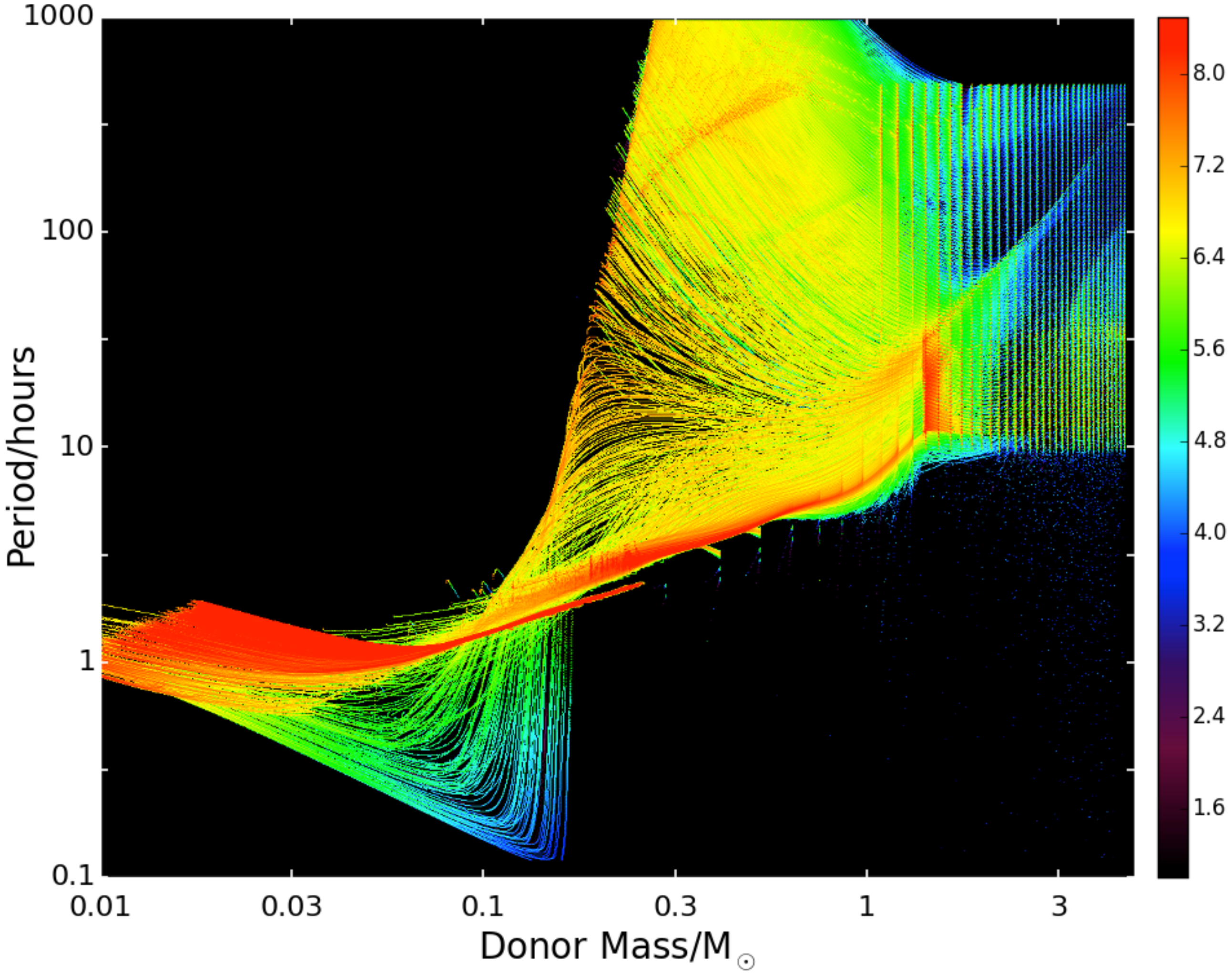}
\caption{56\,000 binary evolution tracks in the plane of orbital period, $P_{\rm orb}$, vs.~the mass of the donor star, $M_{\rm don}$.  In all cases the accreting star is a white dwarf.  The three-dimensional grid of starting models in $P_{\rm orb}$, $M_{\rm don}$, and white dwarf mass, $M_{\rm wd}$, is listed in Table \ref{tbl:initial}. The image is comprised of a discrete grid of $1000 \times 1000$ pixels which is color-coded by the logarithm of the accumulated evolutionary time steps in each pixel for all tracks crossing the array.  Red corresponds to long evolution times, while blue shading represents more rapid portions of the evolution.  The red to blue scaling corresponds to a factor of $\sim$$10^5$ difference in time.}
  \label{fig:time}
\end{figure}

\section{The Grid of Models}
\label{sec:grid}

We have chosen to start our binary evolution models using a grid in three-dimensional parameter space, $M_{\rm wd}$, $M_{\rm don}$, and $P_{\rm orb}$, rather than starting with the output of a population synthesis code (e.g., `BSE'; Hurley, Tout, \& Pols~2002; `StarTrack'; Belczynski et al.~2008) which generates post-common envelope binaries containing a normal star and a white dwarf.  There are two reasons for this choice (as outlined in the Introduction).  First, all such binary population synthesis codes (`BPS'), which start from primordial binaries, have inherent uncertainties, such as the correct parameters to use for the common-envelope phase, and the specific angular momentum carried away by non-conservative, but stable, mass transfer.  Second, the use of an initial regular grid of models ensures that no important evolutionary tracks will be missed because of either biases about what the initial configurations are, or any inaccurate aspects of the BPS codes.

Parameter space for binary evolution studies involving a white dwarf accretor is inherently three-dimensional because of the requirement for choosing three independent system parameters: $M_{\rm wd}$, $M_{\rm don}$, and $P_{\rm orb}$.  This is as opposed to the problem of accreting neutron stars in binaries (see, e.g., Lin et al.~2011) where only $M_{\rm don}$ and $P_{\rm orb}$ need to be chosen (modulo the uncertainty in the neutron star natal mass (see, e.g., Schwab et al.~2010, Valentim et al.~2011; \"Ozel et al.~2012; Kiziltan et al.~2013).  Our starting grid of models for this study is given in Table \ref{tbl:initial}.

\begin{deluxetable}{lcccc}
\centering
\tablecaption{Initial Models of Accreting White Dwarf Binaries}
\tablewidth{0pt}
\tablehead{
\colhead{Parameter} &
\colhead{min value} &
\colhead{max value} &
\colhead{step size}  &
\colhead{\#}
}
\startdata
$M_{\rm don}$ ($M_\odot$)  & 0.3  &  4.7  &  0.11 & 40   \\
$M_{\rm wd}$ ($M_\odot$)    & 0.1 & 1.4 & 0.1 & 14 \\
$P_{\rm orb}$ (hours)  & 10 & 485 & $\delta P/P = 0.04$ & 100\\
\hline
Total models & ... & ... & ... & 56,000\tablenotemark{a}
\enddata
\tablenotetext{a} {\scriptsize See Note Added in Manuscript.}
\label{tbl:initial}
\end{deluxetable}

Overall, this grid is, to the best of our knowledge, at least an order of magnitude larger than anything like it, except for the Chen et al.~(2014) study which had about 50\% the number of evolution tracks.  In the latter study Chen et al.~(2014) focused on SXSs and producing Type Ia SNe, while the more general evolution results were subsumed within their statistical/population study.  The choice of the highest value of $M_{\rm don} = 4.7 \,M_\odot$ is the maximum plausible value for which stable mass transfer onto the maximum mass WD (i.e., $1.4 \,M_\odot$) could occur.

\section{Illustrative Evolution Tracks}
\label{sec:illustrative}

In Fig.~\ref{fig:scheme} we show a set of 18 illustrative evolution tracks in order to set the stage for viewing the larger collection of models that we have generated.  Because most of our results are presented in the $P_{\rm orb}-M_{\rm don}$ plane, we also show this small sample set of tracks in the same plane.  The tracks are color coded according to the various evolutionary branches, described below.  The tracks include an assortment of white-dwarf accretor masses of $M_{\rm wd}$ of 1.4, 1.0, 0,7, and 0.5 $M_\odot$ (not indicated on the figure).

\begin{figure}
\centering
\includegraphics[angle=0, width=\columnwidth]{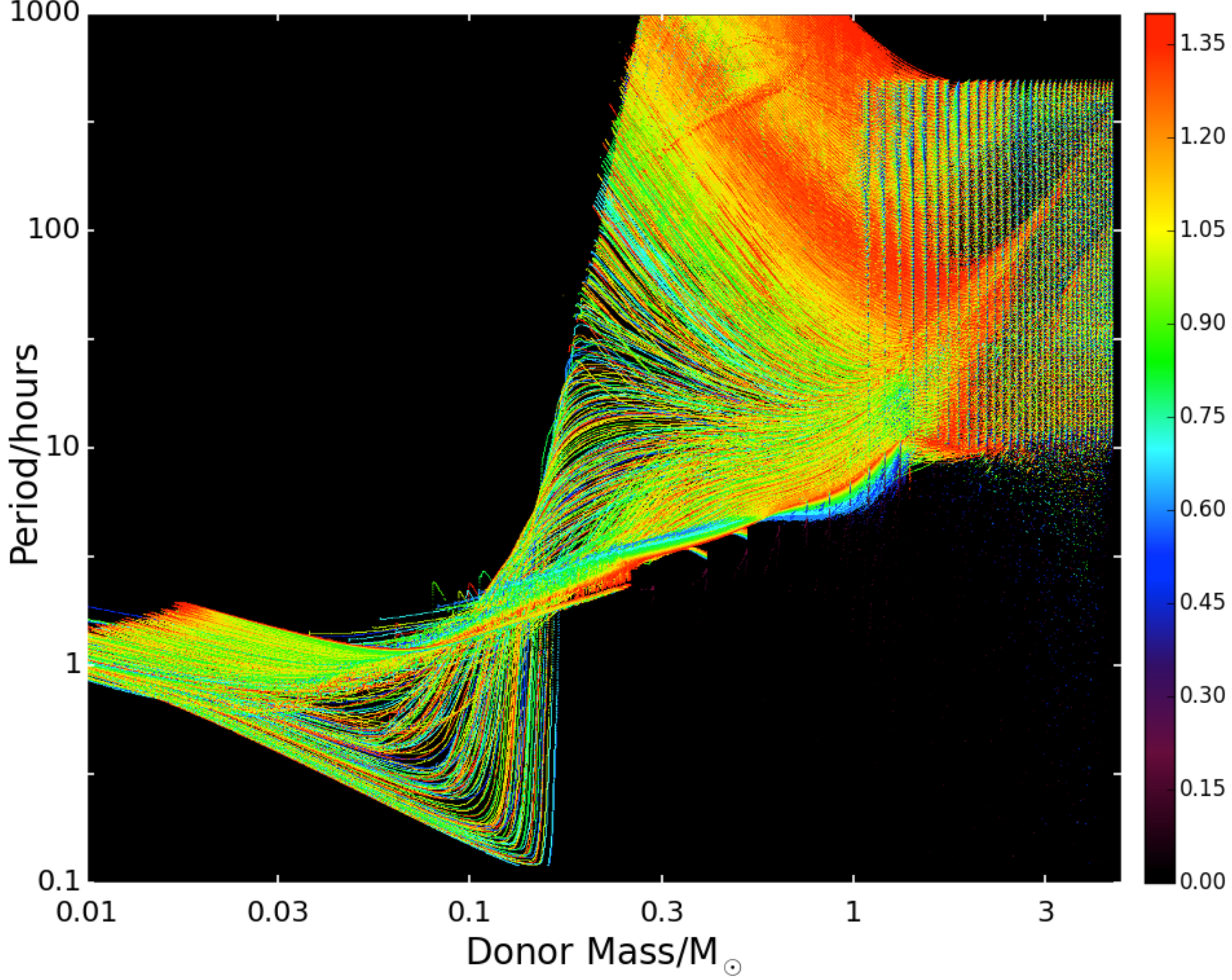}
\caption{Evolution tracks in the $P_{\rm orb}-M_{\rm don}$ plane, color coded by the median value of the mass of the accreting white dwarf.  The color scale is linear in mass with red corresponding to $M_{\rm wd} = 1.2-1.4 \, M_\odot$, while green represents $M_{\rm wd} \simeq 0.80 \,M_\odot$, and blue $\lesssim 0.5 \, M_\odot$.}
 \label{fig:WD}
\end{figure}

\begin{figure}
    \centering
\includegraphics[angle=0, width=1.02\columnwidth]{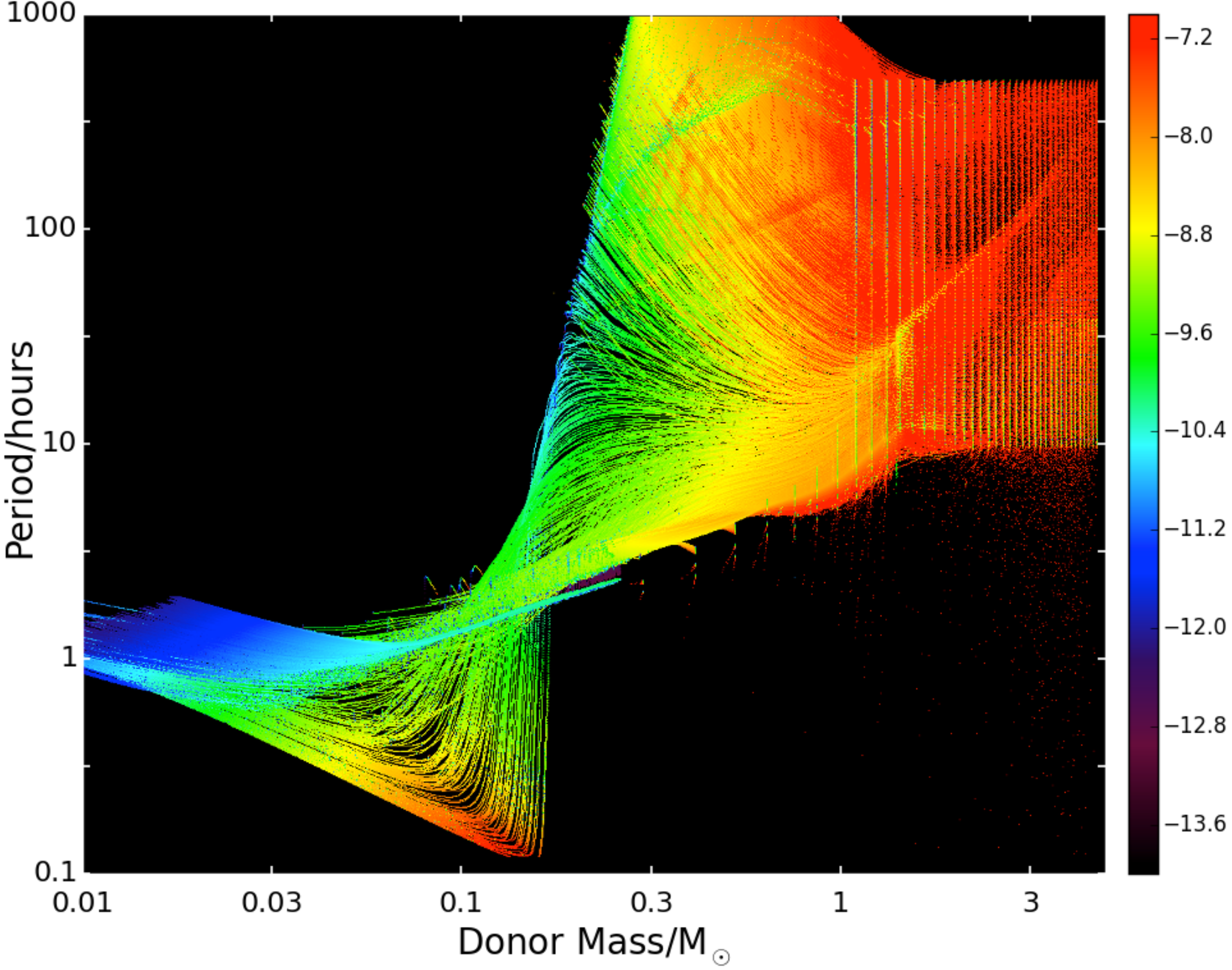}
\caption{Evolution tracks in the $P_{\rm orb}-M_{\rm don}$ plane, color coded by the median value of the mass transfer rate, $\dot M$.  The color scale is logarithmic in $\dot M$ with red corresponding to $\gtrsim 10^{-8} M_\odot$ yr$^{-1}$, while green represents $3 \times 10^{-10} M_\odot$ yr$^{-1}$, and blue $\lesssim 10^{-11} \, M_\odot$ yr$^{-1}$.}
 \label{fig:mdot}
\end{figure}

Four of the tracks in Fig.~\ref{fig:scheme} are labeled as `Cataclysmic Variables' (`CVs'; blue curves), and these represent the evolutionary paths for classical CVs (see, e.g., RVJ, Howell et al.~2001; Goliasch \& Nelson et al.~2015).  These are systems for which mass transfer commences before the core of the donor star can become significantly evolved.  While the donor masses are higher than about 0.25 $M_\odot$ the evolution is presumed to be driven by magnetic braking with typical mass transfer rates of $\dot M \simeq 10^{-9} - 10^{-8} \, M_\odot$ yr$^{-1}$.  Once the donor star becomes completely convective, the magnetic braking is assumed to turn off, or be substantially reduced (RVJ; Spruit \& Ritter 1983).  The donor star, which is then considerably bloated and out of thermal equilibrium, then shrinks inside of its Roche lobe and mass transfer ceases.  This is the conventionally adopted explanation for the so-called `period gap' in CVs over the period range 2-3 hours (these appear as short vertical jogs in the evolution tracks). In addition to the period gap in CVs, further evidence for the disrupted magnetic braking scenario includes: the increased donor star radii above the gap (Knigge 2006); studies of PCEBs found with SDSS in comparison with population models (Schreiber et al.~2010); and recent direct evidence for detached CVs crossing the gap (Zorotovic 2016).  After the orbit is brought back into Roche-lobe contact via gravitational radiation losses, the mass transfer recommences, this time driven largely by gravitational radiation.  Once degeneracy sets in, the donor star does not contract as rapidly due to mass loss, and the orbit starts to widen.  After a Hubble time, the donor stars typically become brown dwarfs with masses of $\approx$20 Jupiter masses.

\begin{figure}
    \centering
    \includegraphics[angle=0, width=\columnwidth]{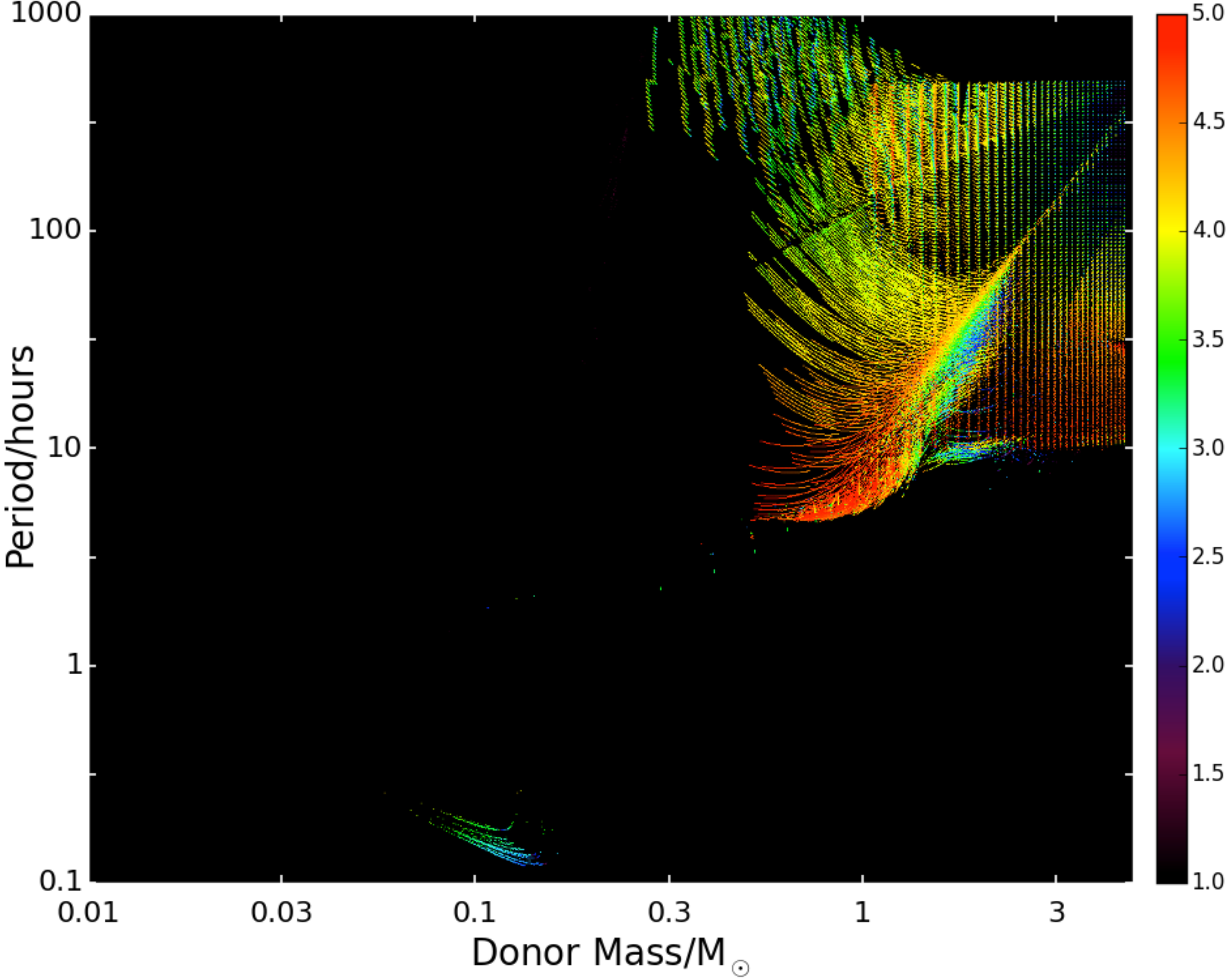}
    \caption{Median evolution dwell times for all systems with accretion rates within the region of stable burning of H on the surface of the white dwarfs.  The full range spans $4 \times 10^{-8} \lesssim \dot M \lesssim 5 \times 10^{-7} \, M_\odot \, {\rm yr}^{-1}$, but depends on the mass of the accreting white dwarf (see text for details). This is the region where SXSs are expected to exist.}
 \label{fig:SXS}
\end{figure}

\begin{figure}
  \centering
\includegraphics[angle=0, width=1.02\columnwidth]{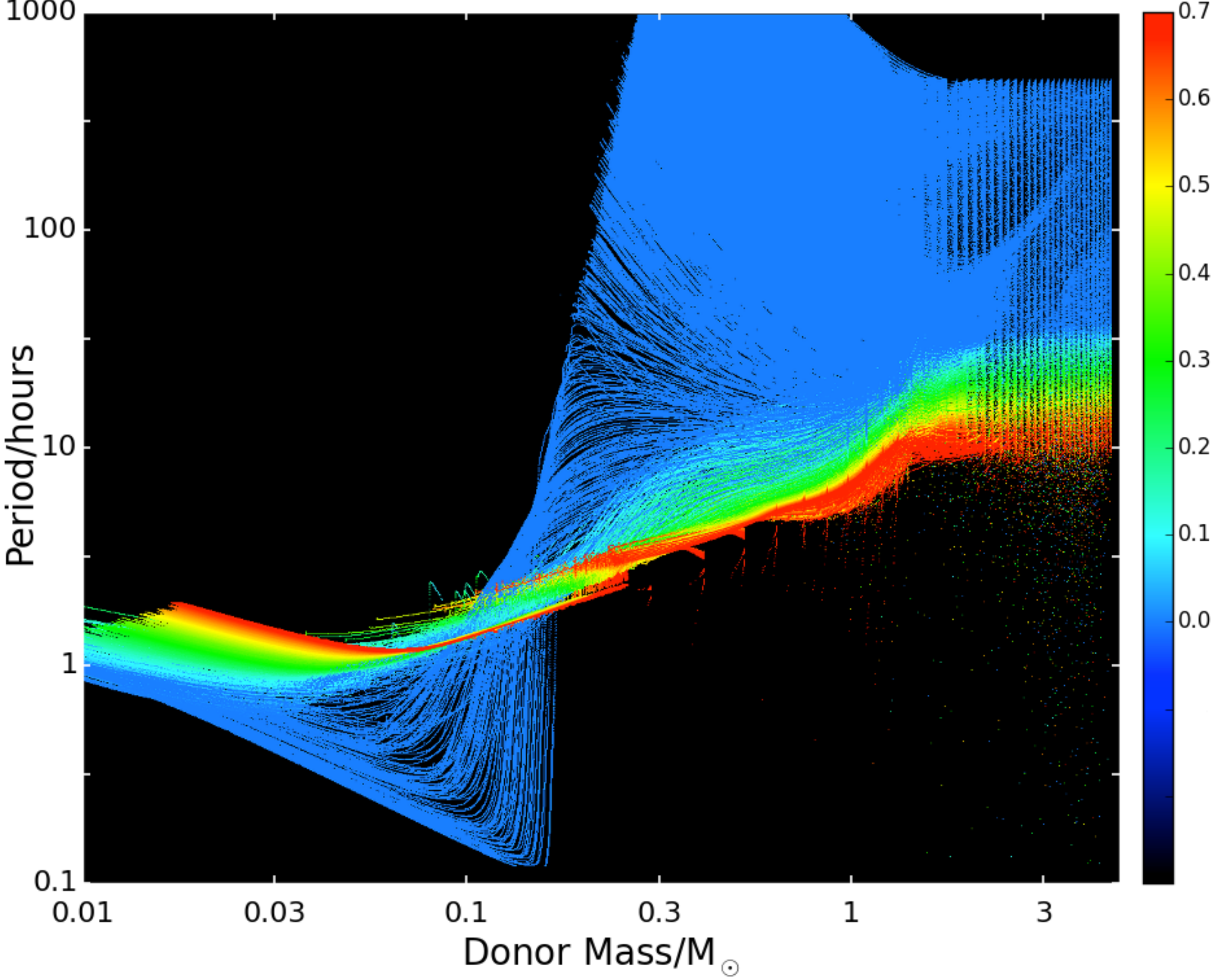}
\caption{Evolution tracks in the $P_{\rm orb}-M_{\rm don}$ plane, color coded by the median value of the hydrogen mass fraction at the center of the donor star.  The color scale is linear in abundance with red corresponding to solar abundance, while green represents $\simeq 40\%$ H by mass, and dark blue $\lesssim 10\%$ H by mass.}
 \label{fig:Hcore}
\end{figure}

There are 9 tracks in Fig.~\ref{fig:scheme} labeled `giants' (red curves), where the core of the donor star is totally hydrogen exhausted, and perhaps a small degenerate He core has formed, at the time that Roche-lobe overflow and mass transfer commence. The orbital period for these tracks typically first decreases and then increases as the donor stars ascend the giant branch.  The initial decrease in orbital period is related to the response of the donor star to thermal timescale mass loss in the early phases of mass transfer (King \& Ritter 1999; Podsiadlowski \& Rappaport 2000).  The thermal timescale mass transfer, in turn, results from the fact that the donor star in most of these cases is more massive than the white dwarf accretor.  Most of these systems have donor stars which evolve up the giant branch with low-mass degenerate He cores ($\approx 0.1-0.2 \, M_\odot$), and their evolution terminates along a well understood curve in the $P_{\rm orb}-M_{\rm don}$ plane (see Rappaport et al.~1995; Lin et al.~2011; Tauris \& van den Heuvel 2014).

Perhaps more interesting than either of the above classes of evolutionary tracks are the three labeled as `ultracompact' systems (purple curves).  These would include AM CVn systems (Podsiadlowski et al.~2003; Nelemans et al.~2010; Kilic et al.~2014).  For these systems the center of the donor star is nearly or completely H-depleted, and they start mass transfer in a narrow orbital period range between those systems that evolve up the giant branch, and those that evolve to become CVs.  This narrow period range is often referred to as the `bifurcation' limit (e.g., Pylyser \& Savonije 1988;  Podsiadlowski et al.~2003; Nelson et al.~2004b; Lin et al.~2011; Lin 2011).   For the input physics assumed in this work, the bifurcation period can be approximately expressed as:
\begin{equation}
P_{\rm orb} \simeq 18.6 + 3.7\,M_{\rm don}/M_\odot ~{\rm hours}
\label{eqn:bifur}
\end{equation}
where $P_{\rm orb}$ is the orbital period at the onset of mass transfer.\footnote{It should be noted that the bifurcation limit is extremely sensitive to the assumed input physics. However, its general behavior does not change.}
With He cores that are partially degenerate, these donors lose enough of their H-rich envelopes to prevent H-shell burning, but have core masses that are too low to allow for He burning.  Therefore, their radii shrink dramatically so that binary periods as short as $\sim$6 minutes can be reached.  These systems are not likely to be plentiful since, for a given donor mass, the range of initial periods that will lead to this state is narrow, perhaps $\lesssim 1$ hour for orbital periods near 25 hours.

Two of the tracks in Fig.~\ref{fig:scheme} (green curves) start to evolve toward the ultracompact phase, but then the donor stars lose all of their envelopes while the orbital period is still at $\sim$10-20 hours.   Because these orbital periods are too long for gravitational radiation to bring the donor star back into Roche-lobe contact, the period is simply frozen at this value.  We refer to these as `transition systems'.

\begin{figure}
    \centering
 \includegraphics[angle=0, width=1.02\columnwidth]{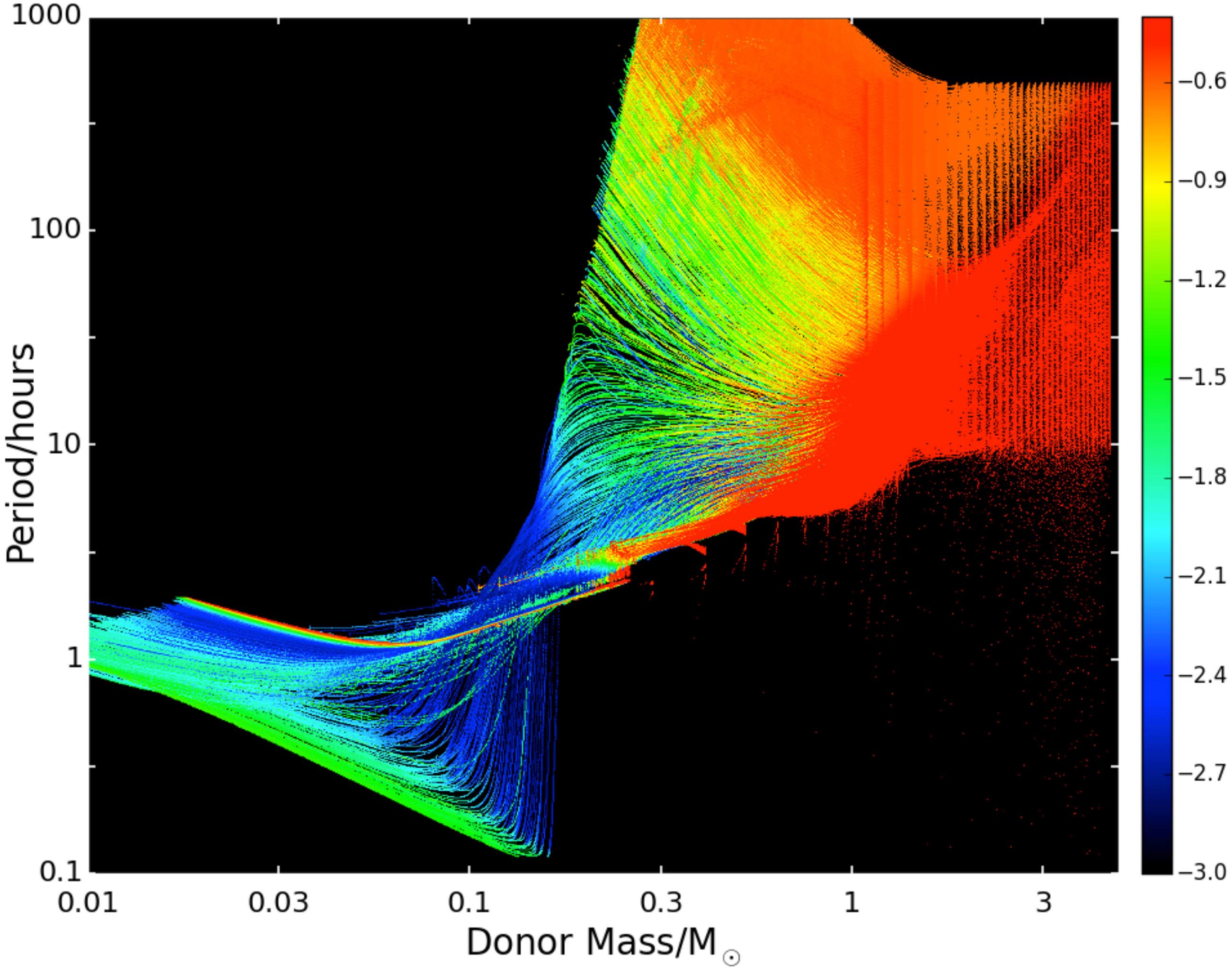}
\caption{Evolution tracks in the $P_{\rm orb}-M_{\rm don}$ plane, color coded by the median value of the C/O ratio at the surface of the donor star.  The color scale is logarithmic in the ratio with red corresponding to C/O = 0.37, while green represents C/O in the range 0.016-0.063, and blue C/O = 0.003-0.008.  For reference, the initial ratio of C/O in the donor star is 0.37. }
 \label{fig:CO}
\end{figure}

In general the stability of mass transfer depends on the mass ratio $M_{\rm don}/M_{\rm wd}$ and the evolutionary state of the donor at the start of mass transfer.  The larger this mass ratio, the more unstable the mass transfer is likely to be for a given internal chemical profile of the donor (i.e., for a specific evolutionary state).  However, as has been found in earlier studies (Rappaport et al.~1994; Di Stefano \& Nelson 1996; Podsiadlowski \& Rappaport 2000), initial mass ratios as high as 3:1 may still be stable, but generally only after a phase of rapid thermal timescale mass transfer (`TTMT') that brings the mass ratio of the system closer to unity. For some recent applications of the mass-transfer stability criterion with respect to the merger of CVs containing helium WD accretors, see Schreiber et al.~(2016) and Nelemans et al.~(2016). We have found that MESA is quite robust at following stable mass transfer through the thermal timescale mass-transfer phase.  However, with MESA, the user needs to set a maximum value of $\dot M_{\rm max}$ above which the code will restrict the mass-loss rate to the upper limit that has been specified.  In some cases, if $\dot M_{\rm max}$ is set sufficiently high, the evolution will still succeed even if hydrodynamic effects are not properly taken into account (e.g., high-velocity winds, mergers).  This causes MESA to take off mass more slowly than the thermal timescale would dictate.  We have therefore set a maximum rate, $\dot M_{\rm max}$, beyond which we terminate the evolution, even if MESA would allow it.

As already mentioned, when the donor star is more massive than the accretor and has a radiative envelope, the mass transfer rate tends to proceed on a thermal timescale (Rappaport et al.~1994; Di Stefano \& Nelson  1996; Podsiadlowski \& Rappaport 2000).  An estimate of this rate is:
\begin{equation}
\dot M_{\rm therm} \simeq f \frac{M_{\rm don}}{\tau_{\rm KH}}
\end{equation}
where $f$ is an adjustable multiplicative factor, and $\tau_{\rm KH}$ is the Kelvin-Helmholtz (`KH') timescale for the entire star, which, in turn, can be estimated by $\tau_{\rm KH} \simeq GM_{\rm don}^2/(R_{\rm don} L_{\rm don})$.  However, the KH time of the outer envelope of the star which is directly involved in the mass transfer may be much shorter than this.  Accordingly, we adopt a factor of $f \sim$300 to take into account the fact that the thermal timescale of the relevant portions of the donor star may be much shorter than $\tau_{\rm KH}$.  Finally, we adopted the following for the maximum allowed $\dot M$ before we terminate the evolution:
\begin{equation}
\dot M \simeq 10^{-5} \frac{\mathcal{R}_{\rm don} \mathcal{L}_{\rm don}}{\mathcal{M}_{\rm don}} M_\odot \,{\rm yr}^{-1}
\end{equation}
where the donor mass, radius, and luminosity are expressed in solar units.

\section{Basic Results in the $P_{\rm orb}-M_{\rm don}$ Plane}
\label{sec:results}

We now present the results of our evolution of 56,000 binaries with accreting white dwarfs.  Most of the results are presented in the form of `images' in the $P_{\rm orb}-M_{\rm don}$ plane where the color represents the amplitude of a `third parameter'.  These images each contain $1000 \times 1000$ discrete pixels into which we map the evolution tracks.  The `third parameter' includes the evolution time, the mass of the white-dwarf accretor, the mass-transfer rate, and the central and surface chemical compositions of the donor star.

Figure \ref{fig:time} shows the accumulated evolutionary dwell time per pixel.  According to the color scale, these range from $10^3-3 \times 10^8$ years. But, since each pixel is only a small portion of the evolution track, the total times involved are as long as $10^{10}$ years to reach the end of the classical CV tracks to as short as $10^6$ years for the ultra-compact sources.  In fact all of the types of evolutionary tracks discussed in connection with Fig.~\ref{fig:scheme} are seen here in a nearly continuous blend.  There are a few things of special note to mention with respect to this figure. First, the location of the grid of initial models is quite evident. The three basic evolutionary branches, compact CVs, giant donors, and ultracompact systems are all clearly included in the figure.  The red region toward very low donor masses along the CV tracks is quite apparent; these evolve very slowly since the only driver of the orbital evolution is gravitational radiation which becomes progressively less effective as the donor mass shrinks.  Interestingly, another region in the $P_{\rm orb}-M_{\rm don}$ plane with slowly evolving systems has $10 \lesssim P_{\rm orb} \lesssim 40$ hours, and $M_{\rm don} \simeq 0.2 \,M_\odot$.  Ironically, these are the same systems which `dive' to ultrashort orbital periods and then evolve very rapidly (the blue region of the ultracompacts).  Systems with donors on the giant branch largely terminate their evolution along the sloping line with a well-defined relation: $P_{\rm orb} \appropto M_{\rm don}^9$ (see Sect.~\ref{sec:final}).

\begin{figure}
    \centering
\includegraphics[angle=0, width=1.02\columnwidth]{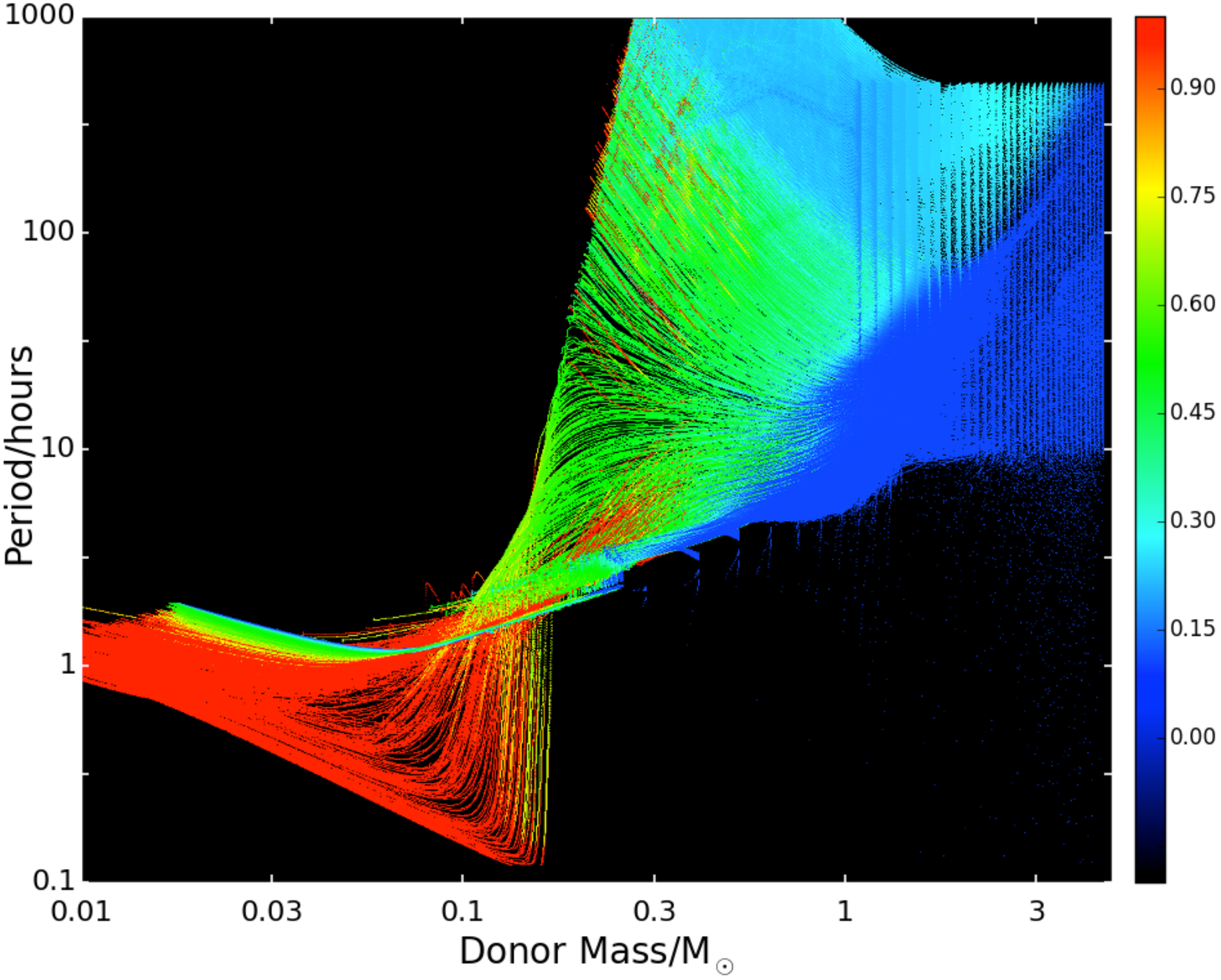}
\caption{Evolution tracks in the $P_{\rm orb}-M_{\rm don}$ plane, color coded by the median value of the N/O ratio at the surface of the donor star.  The color scale is linear in the ratio with red corresponding to N/O $\gtrsim $ unity, while green represents N/O $\simeq 0.5$, and blue is N/O $\simeq 0.1$.  For reference, the initial ratio of N/O in the donor star is 0.11.}
 \label{fig:NO}
\end{figure}

\begin{figure}
    \centering
\includegraphics[angle=0, width=1.02\columnwidth]{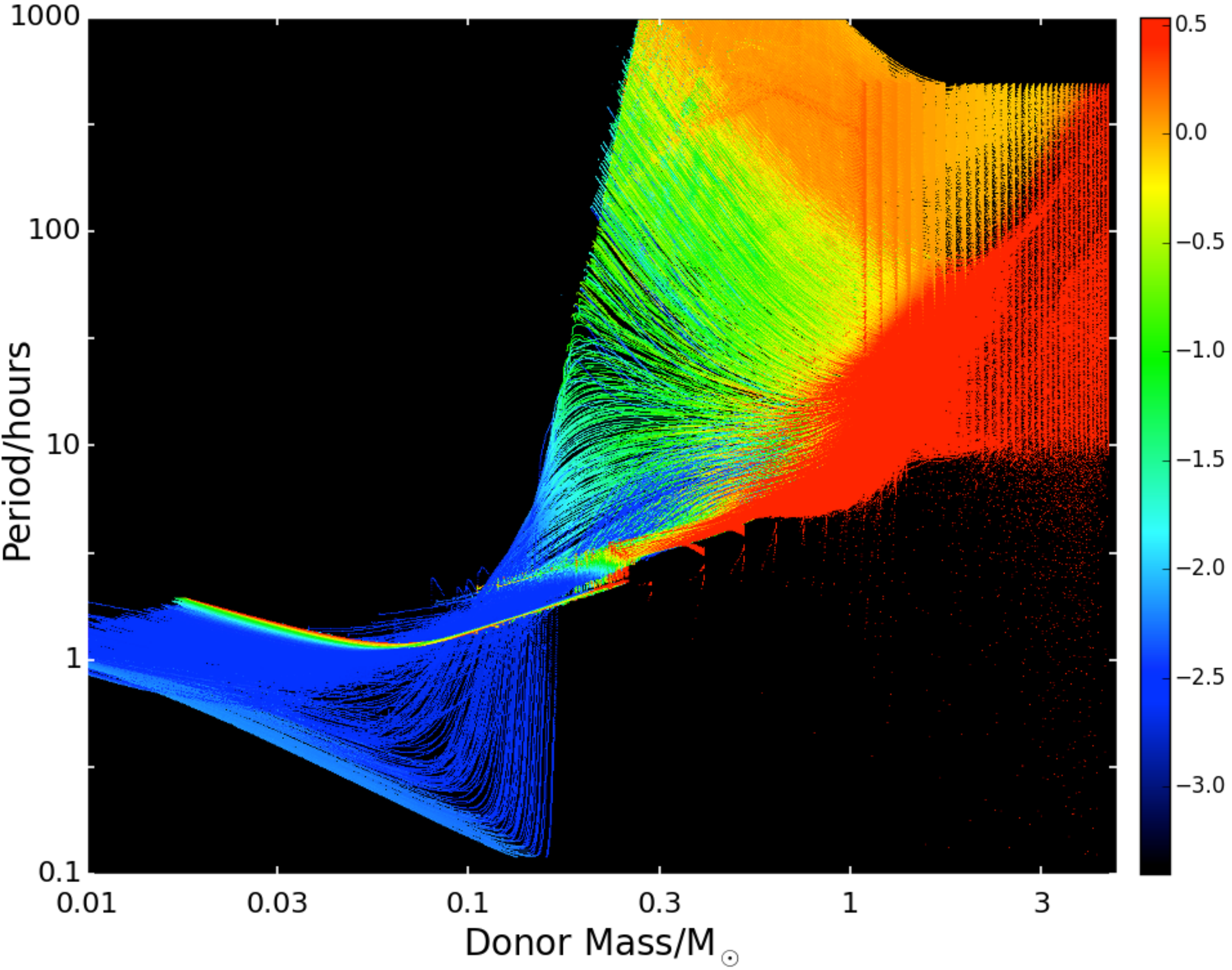}
\caption{Evolution tracks in the $P_{\rm orb}-M_{\rm don}$ plane, color coded by the median value of the C/N ratio at the surface of the donor star.  The color scale is logarithmic in the ratio with red corresponding to C/N = 3.4, while green represents C/N in the range 0.04-0.3, and blue is C/N $\lesssim 0.006$.  For reference, the initial ratio of C/N in the donor star is 3.4.}
 \label{fig:CN}
\end{figure}

Figure \ref{fig:WD} shows the $P_{\rm orb}-M_{\rm don}$ plane color coded according the median mass of the white-dwarf accretor.  Most of the regions are greenish in color with sprinklings of red and blue showing through.  This results from the fact that most masses for the white dwarf accretors can end up in most parts of the diagram, and therefore the median value corresponds to roughly 0.8 $M_\odot$.  The only exception is the red region toward higher donor masses and longer orbital periods.  These regions are only accessible starting with the more massive donor stars, which, in turn, require the most massive accreting white dwarfs in order to maintain stability during the mass transfer.

Figure \ref{fig:mdot} shows the $P_{\rm orb}-M_{\rm don}$ plane color-coded by the logarithm of the median value of the mass-transfer rate.  In general, we can see that systems with large initial donor masses have high $\dot M$'s (exceeding $10^{-6} \, M_\odot$ yr$^{-1})$, while, by contrast, old CVs with short orbital periods have very low $\dot M$'s ($10^{-10}-10^{-11} \, M_\odot$ yr$^{-1})$.  Also note that the ultracompact systems have relatively high values of $\dot M$ as well (at $\sim$$10^{-9} \, M_\odot$ yr$^{-1}$).

In order to have covered some 7 orders of magnitude in $\dot M$ in Fig.~\ref{fig:mdot} we had to sacrifice good `color resolution', and even systems with a factor of 10 difference in $\dot M$ may appear with the same color.   Therefore, in order to bring out an important and interesting facet of this diagram, we show another version of this figure, but this time restricted to systems where stable hydrogen burning can take place on the surfaces of the white dwarfs (Iben 1982; Nomoto 1982, 1988; Iben \& Tutukov 1989).  Analytic expressions for the upper and lower boundaries of this regime, as functions of the white dwarf mass, are given by Eqns.~(3) and (4) of Chen et al.~(2014) as approximations to the numerical results of Iben \& Tutukov (1989).  Over all white dwarf masses of interest, the critical values of $\dot M$ fall roughly over the range of $4 \times 10^{-8} - 5 \times 10^{-7} M_\odot$ yr$^{-1}$.  The image indicating where the stable surface  H-burning white dwarfs lie in the $P_{\rm orb}-M_{\rm don}$ is shown in Fig.~\ref{fig:SXS} and is color-coded by the median evolutionary dwell time in a given pixel.  The reason why this particular region is of interest is that this is where one might expect to find most supersoft X-ray sources.  The red and yellow regions spanning $0.6 \lesssim M_{\rm don} \lesssim 2 M_\odot$ and orbital periods of $\approx 5-100$ hours, should contain many of the SXS.  In fact, CAL 83, CAL 87, and RX J0019.8+2156 (Odendaal et al.~2015; Ablimit \& Li~2015; Becker et al.~1998) are in this region (see also Rappaport et al.~1994 and Chen et al.~2014).  Note, however, that the shortest-period UCs that appear faintly in this plot are not likely to be SXSs because most of the matter being transferred to the WD is He, not H.  Because the `critical' mass-transfer rates for SXSs scale roughly as $1/X$, where $X$ is the H mass fraction, we would expect the critical steady-burning rates to increase by a factor of $\sim$100 or more.Ê The critical values of $\dot M$ for a UC to become a SXS are so high that they are simply not met.

 We next explore the H mass fraction at the center of the donor stars which is shown in Fig.~\ref{fig:Hcore}.  What we can see clearly is that the classical H-rich CVs, with minimum periods of $\sim$70 minutes, remain H rich down to the oldest ages and lowest donor masses. By contrast, directly below these CV tracks one can see an impressive sequence of systems that are more and more H depleted and have shorter and shorter minimum orbital periods. The donor stars in these systems were progressively more evolved at the onset of mass transfer.  In turn these tracks blend smoothly into the evolution tracks of the ultracompact systems. The systems with donors which evolve up the giant branch, naturally, are H-exhausted in their cores. Also, as expected, the donors in the ultracompact systems have virtually no H in their cores. Otherwise, orbital periods as short as $5- 20$ minutes would not be possible.

Finally, we look at the surface chemical composition of the donor stars.  In contrast to the central chemical composition, the surface composition might be inferred either via spectral features of the donor star or from an accretion disk around the accreting white dwarf (see, e.g., Hamilton et al.~2011).  In Figs.~\ref{fig:CO}, \ref{fig:NO}, and \ref{fig:CN} we show the median values of the surface composition ratios of C/O, N/O, and C/N, respectively.  Perhaps the most dramatic of these is Fig.~\ref{fig:CN} showing the C to N ratio.  The initial (i.e., primordial) mass ratios of O:C:N as inputs to MESA were 1.0:0.37:0.11.  As the surface composition plots show, these ratios evolve by large factors, depending on where in the $P_{\rm orb}-M_{\rm don}$ plane a system is found.  Some illustrative quantitative values of the ratios C/N, N/O, and C/O are listed in Table \ref{tbl:surf_comp} for five locations in the evolution diagrams representing different types of WD-accreting binaries: (1) donors on the giant branch; (2) `normal' CVs; (3), CVs with significantly evolved donor stars; (4) transition systems with $P_{\rm orb} \simeq 10$ hr and $M_{\rm don} \simeq 0.2 \, M_\odot$; and (5) ultracompact binaries.  As can be seen from Table \ref{tbl:surf_comp} and Figs.~\ref{fig:CO}--\ref{fig:CN}, the CNO ratios have evolved from their primordial composition by factors of a few for donor stars on the giant branch, to factors of $\gtrsim 100$ for the ultracompact binaries.  The sense of these composition changes is that N becomes greatly enhanced at the expense of C, while O also becomes modestly depleted (see, e.g., Clayton 1968; Nelemans et al.~2010; Kochanek 2016).  This is a signature of the C abundance quickly coming into equilibrium (at very low levels), whereas O has not come into nuclear quasi-equilibrium because the interior temperatures are too low.

The changes in the surface chemical abundances result largely from the CNO bi-cycle. The limiting (i.e., the slowest) reaction in the CN portion of the cycle is ${\rm N}^{14}(p,\gamma){\rm O}^{15}$ and causes ${\rm N}^{14}$ to build up and ${\rm C}^{12}$ to become relatively depleted.  The branching ratio pertinent to the NO portion of the cycle is more than three orders of magnitude smaller than for the CN portion and does not play an important role in the replenishment of ${\rm C}^{12}$ to complete the second loop in the cycle. The reactions ${\rm O}^{16}(p,\gamma){\rm F}^{17}$ followed by the inverse beta-decay of F$^{17}$ and ${\rm O}^{17}(p,\alpha){\rm N}^{14}$ serve to deplete the primordial O in favor of N, but only on much longer timescales.  In the low-mass donors ($\lesssim 1\,  M_\odot$) the CNO cycle plays little role in H-burning, but of course these are the same donor stars that experience little chemical evolution during their lifetimes in CV systems. For stars more massive than the Sun, especially our donor stars with masses in the range of $1.5-4 \, M_\odot$, the CNO process becomes the dominant mode for consuming H, and they are the stars whose descendants show the greatest degree of chemical evolution and isotopic ratio shifts.  For CVs late in their evolution, the transferred mass comes directly from the partially degenerate remnant cores of the donor stars, and this is the source of the nitrogen-enhanced material.  For UC systems, after a large amount of mass is lost, nitrogen enhancement can be extreme as the outer layers of the donor reach the core where CNO burning had dominated.  In the case of giant donor stars with highly degenerate cores, the nuclear source powering these systems up the giant branch is H-shell burning, which again is dominated by the CNO cycle.  Some of that nitrogen-enriched material has been dredged up via convective mixing in the giant's envelope and is transferred to the accretor.  Thus, for many CV-like systems, it is possible in principle, to observe very large attenuations in the C/N ratio in the material that is deposited in the accretion disk surrounding the WD accretor for certain phases of a CV's evolution.

\begin{deluxetable}{lccc}
\centering
\tablecaption{Surface Composition}
\tablewidth{0pt}
\tablehead{
\colhead{Evol.~Phase} &
\colhead{C/N} &
\colhead{N/O} &
\colhead{C/O}
}

\startdata
Primordial\tablenotemark{a}  &  3.4 & 0.108 & 0.37 \\
Giants\tablenotemark{b} &  $1-1.3$  & 0.23 & 0.27 \\
CVs\tablenotemark{c} &  $0.22-3$  & $0.11-0.3$ & $0.11-0.33$ \\
Evol.~CVs\tablenotemark{d} &  $0.003-0.03$ & $0.3-1$ & 0.$003-0.03$ \\
Transition\tablenotemark{e} &  0.07 & 0.53 & 0.03 \\
UCs\tablenotemark{f} &  $0.003-0.006$ & $0.7-5$ & $0.003-0.04$\\
\enddata
\tablenotetext{a}{MESA input.}
\tablenotetext{b}{$M_{\rm don} \lesssim 1 \,M_\odot$ and $P_{\rm orb} > 100$ hrs.}
\tablenotetext{c}{Longest period CVs at any given donor mass.}
\tablenotetext{d}{CVs with periods down to 40 minutes.}
\tablenotetext{e}{Systems with $P_{\rm orb} \simeq 10$ hr, and $M_\odot \simeq 0.2 \,M_\odot$.}
\tablenotetext{f}{Ultracompact systems with $P_{\rm orb} \simeq 10$ minutes.}
\label{tbl:surf_comp}
\end{deluxetable}

\section{Orbital Period Minima}
\label{sec:Pmin}

\subsection{CV minimum period vs. He content}

\begin{figure}
    \centering
    \includegraphics[angle=0, width=\columnwidth]{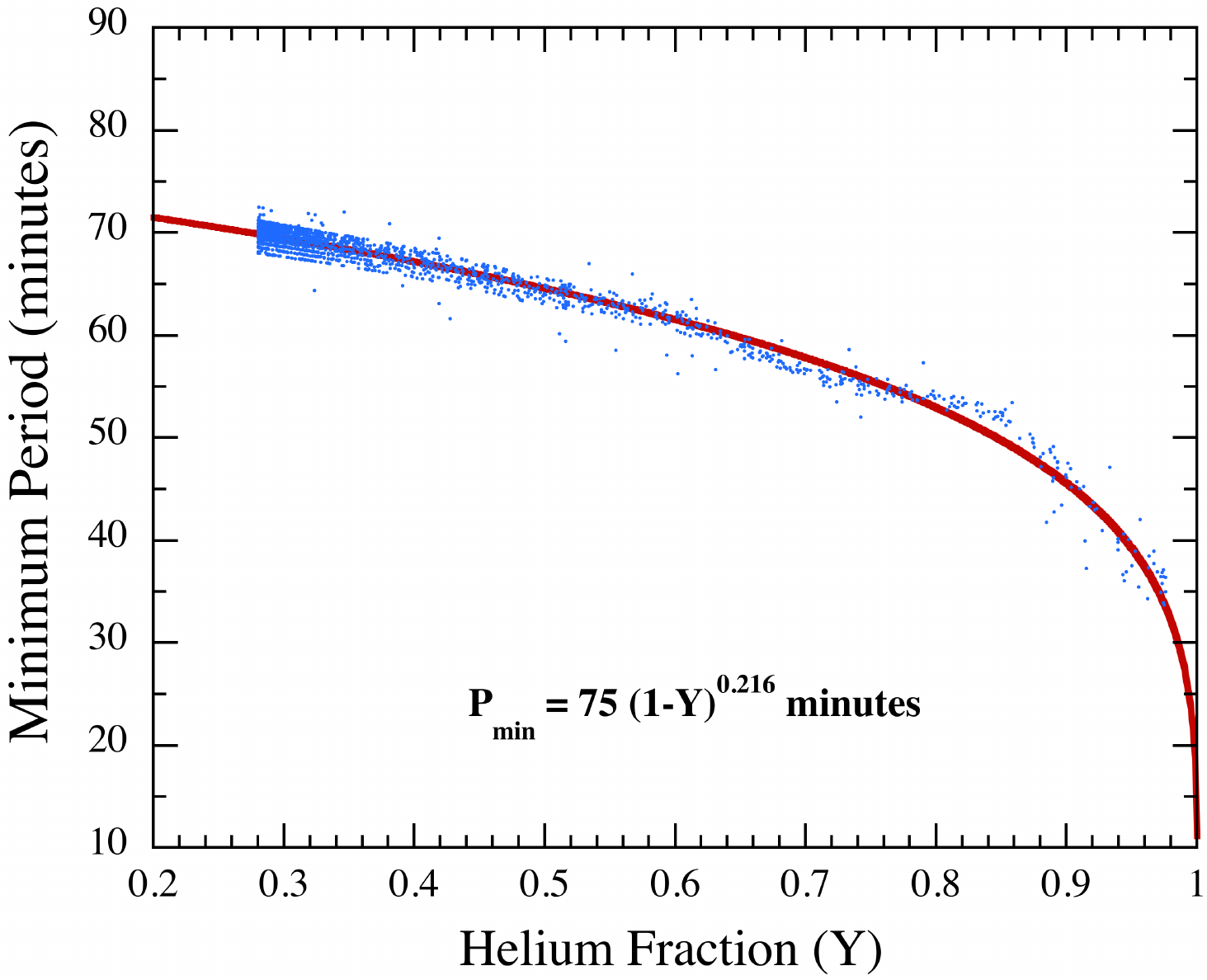}
     \includegraphics[angle=0, width=\columnwidth]{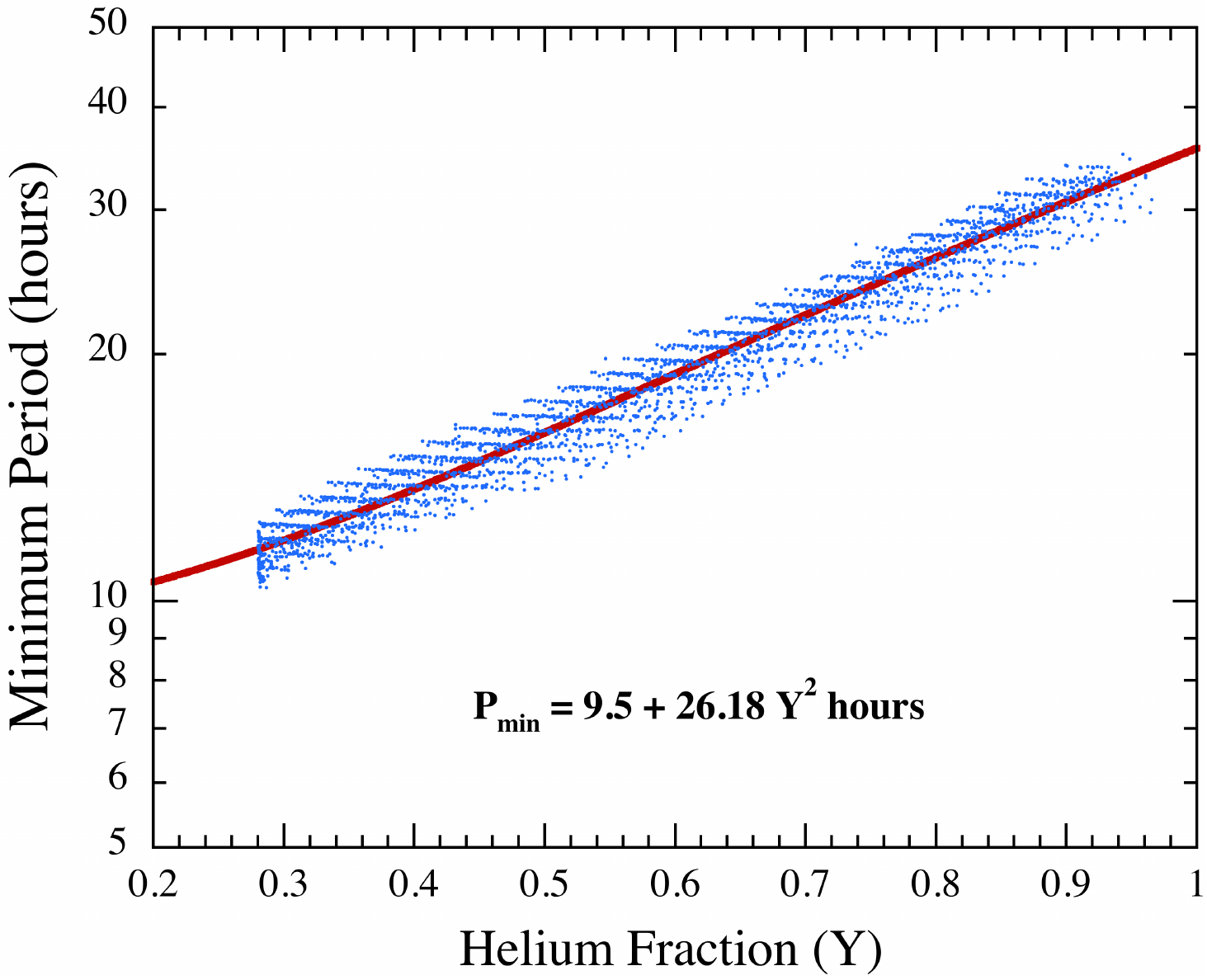}
    \caption{Orbital period minima vs.~the He content in the cores of the donor stars.  The top panel is for cataclysmic variables, while the bottom panel is for the thermal-timescale mass-transfer portion of the evolution.}
  \label{fig:PminY}
\end{figure}

We have collected from our evolution tracks the minimum orbital periods of systems on the CV branch as a function of the central He fraction of the donor star.  The results are plotted in the top panel of Figure \ref{fig:PminY}.  For normal solar abundances, the minimum orbital period has the well-known conventional value of $\sim$70 minutes (see, e.g., RJW).  And, as the plot shows, as the central He fraction approaches unity, the minimum orbital period dives to values as low as $\sim$10 minutes.  We have fit a simple analytic function to this curve which is accurate to $\sim$1-2 minutes:
\begin{equation}
P_{\rm orb, min} \simeq 75 \,(1-Y)^{0.216} ~{\rm minutes}
\label{eqn:pmin1}
\end{equation}
We can explain the trend of this result by first noting that the donor stars near the minimum period on the CV branch have substantially degenerate interiors.  We also know that for cold electron degenerate stars, in the relevant mass range, their radius depends on mass as
\begin{equation}
R_{\rm don} \propto (1+X)^{5/3} M_{\rm don}^{-1/3}
\end{equation}
(see, e.g., Nelson \& Rappaport 2003). Since the orbital period of a binary with a donor star filling its Roche lobe is approximately proportional to $R_{\rm don}^{3/2} M_{\rm don}^{-1/2}$, this implies that at minimum orbital period:
\begin{equation}
P_{\rm orb,min} \propto (1+X)^{5/2} M_{\rm don}^{-1}
\label{eqn:pmin2}
\end{equation}
It is less obvious how the mass of a degenerate donor star at minimum orbital period depends on its composition; however, we have extracted this dependence from the current calculations.  If we fit that dependence to a simple function of $Y$, we find
\begin{equation}
M_{\rm don, min} \simeq 0.0878 - 0.09Y+0.032 Y^2 ~.
\label{eqn:mmin}
\end{equation}
Then we can show quasi-analytically from Eqns.~(\ref{eqn:pmin2}) and (\ref{eqn:mmin}) that the dependence on $Y$ exhibited by Eqn.~(\ref{eqn:pmin1}) and the top panel of Fig.~\ref{fig:PminY} is quite reasonably to be expected.  However, it remains to be shown analytically why the relation in Eqn.~(\ref{eqn:mmin}) holds at the minimum orbital period.

\subsection{Minimum Period for Ultracompact Systems}
\label{sec:PminUC}

As we showed in the previous section, for even moderately degenerate donor stars at orbital period minimum, the period is related to the donor mass and its H fraction by:
\begin{equation}
P_{\rm orb,min} \propto (1+X)^{5/2} M_{\rm don}^{-1}~ \propto ~M_{\rm don}^{-1}
\label{eqn:PminUC}
\end{equation}
However, we have found that all the ultracompact systems have X = 0 in the center of their cores.  Thus, the expression reduces to the rightmost term in Eqn.~(\ref{eqn:PminUC}).

\begin{figure}
    \centering
    \includegraphics[angle=0, width=\columnwidth]{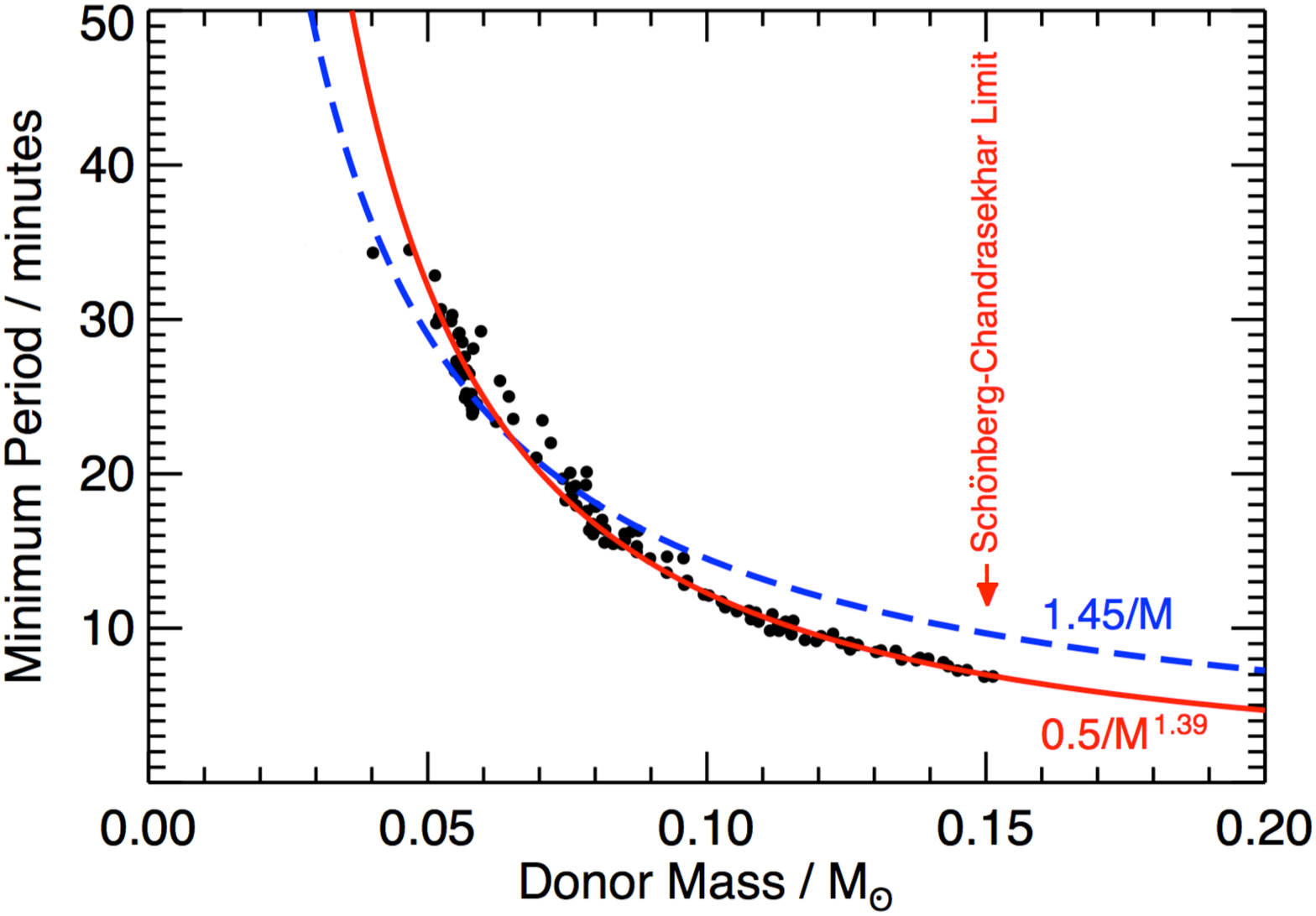}
    \caption{Orbital period minima vs.~the mass of the donor star for ultracompact systems.  The smooth curves are fits to functions of the form $P_{\rm orb,min}= {\rm constant}/M_{\rm don, min}^{\alpha}$ where $\alpha$ is fixed at unity for the dashed blue curve, but has a value of 1.39 for the red curve.}
  \label{fig:PminUC}
\end{figure}

We plot in Fig.~\ref{fig:PminUC} the mimimum orbital period for UC systems against the mass of the donor star at $P_{\rm orb,min}$.  We show for reference a function of the form $P_{\rm orb,min} = c/M_{\rm don,min}$ where $c$ is an adjustable parameter.  While the fit is not very tight, it does show that this very simple expression (based on Eqn.~\ref{eqn:PminUC}) is suggestively on the right track.  We also include a fit of the more general form $P_{\rm orb,min} = c/M_{\rm don,min}^{\alpha}$ where $\alpha$ is a second adjustable parameter, and that fits quite well.

Finally, in this regard we note that the highest mass donor star found in our model UC systems is at $0.15 \,M_\odot$.  This is close to, and suggestive of, the Sch\"onberg-Chandrasekhar (`S-C') limiting mass of $\sim$8\% of the initial stellar mass of the donor.  Most of the UC systems have initial donor-star masses of $1.1-1.6 \, M_\odot$ with corresponding S-C masses of $0.09-0.13 \, M_\odot$, above which the system would have evolved up the giant branch.

\subsection{Minimum Period During Thermal Timescale Mass Transfer}
\label{sec:PminTTMT}

Systems initially undergoing thermal-timescale mass transfer also exhibit a minimum in their orbital period, provided that the evolution does not end in a dynamical instability.  The minimum orbital period during thermal timescale mass transfer is plotted in the bottom panel of Fig.~\ref{fig:PminY}.  These points can be fit to a very simple function of the form:
\begin{equation}
P_{\rm orb, min} \simeq 9.5 + 26.2 \,Y^2 ~{\rm hours}
\label{eqn:pmin3}
\end{equation}

For binary systems where the mass transfer is 100\% non-conservative as we assume here, and is certainly the case during at least most of the TTMT  phase, we can compute analytically the orbital period after any amount of the donor star has been transferred (and ejected from the system).  The ratio of $P_{\rm orb}$ to the initial value, $P_{\rm orb, i}$, is given by
\begin{equation}
\frac{P_{\rm orb}}{P_{\rm orb, i}} = e^{3(M_{\rm 2} - M_{\rm 2,i})/M_{\rm wd}}  \left(\frac{M_{2i}}{M_2}\right)^3  \left( \frac{M_{\rm wd} + M_{2i}} {M_{\rm wd} + M_2} \right)^2
\end{equation}
where $M_{2,i}$ and $M_2$ are the donor masses initially and after some mass loss, respectively.  We can find the minimum of this ratio by differentiating with respect to $M_2$ and setting the result equal to 0.  This yields
\begin{equation}
M_{\rm 2,min} = \frac{1}{3} \left[1+\sqrt{10} \right] M_{\rm wd} ~=~ 1.387 M_{\rm wd}
\end{equation}
White dwarfs that can survive TTMT tend to be among the most massive, with $1 \, M_\odot \lesssim M_1 < 1.4 \,M_\odot$.  This implies that the donor stars at minimum orbital period have masses in the range of $1.4 \, M_\odot \lesssim M_2 \lesssim 2 \, M_\odot$.  At this time, we do not have an analytic expression for the radii of such stripped stars as a function of their central helium fraction.  Therefore, we are not yet able to complete the analytic derivation of the relation given by Eqn.~(\ref{eqn:pmin3}) and shown in the bottom panel of Fig.~\ref{fig:PminY}. We leave that as an exercise for future studies.

\section{Final States of the Systems}
\label{sec:final}

\begin{figure}
    \centering
    \includegraphics[angle=0, width=\columnwidth]{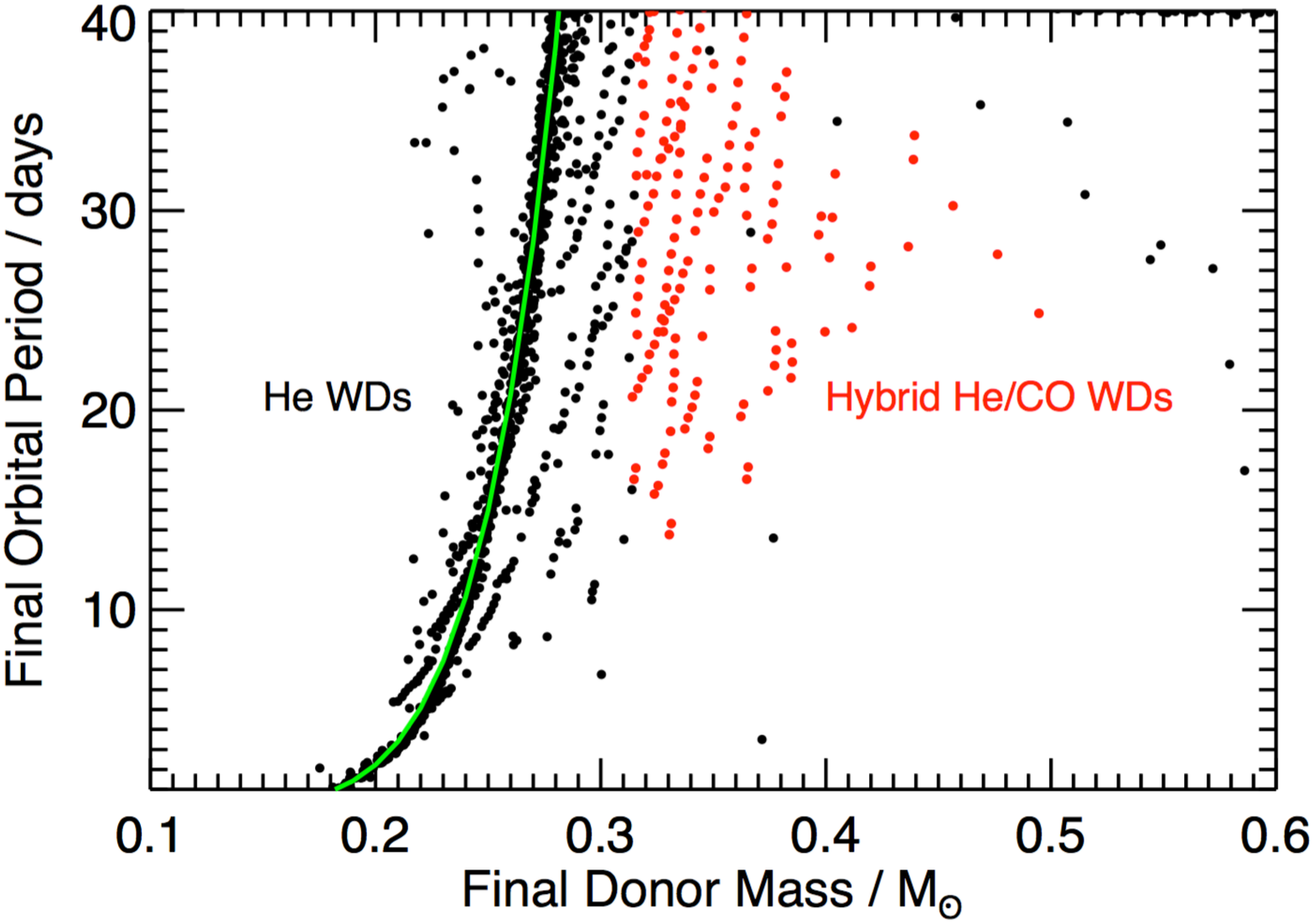}
    \caption{The final orbital period vs.~the final mass of the donor star, i.e., essentially the mass of its stripped degenerate core.  The white dwarf relics of the donor stars (seen on this plot) with masses $\lesssim 0.32 \, M_\odot$ are pure He in composition (color-coded in black) and result from donor stars of initial mass $\lesssim 2.2 \, M_\odot$.  The lime-green curve that goes through the left edge of this band is the theoretical $P_{\rm orb}-M_{\rm wd}$ relation taken from Lin et al.~(2011).  By contrast, the white dwarf relics of the donor star with masses $\gtrsim 0.35 \, M_\odot$ are hybrid He/CO WDs (color-coded in red) and result from donor stars of initial mass $\gtrsim 2.2 \, M_\odot$.}\label{fig:fig10}
\end{figure}

We have collected all of the final donor masses for systems with final orbital periods greater than 1 day.  These are systems where some significant chemical evolution has taken place in the cores of the donor stars, and they evolve at least somewhat up the giant branch.  A plot of the final orbital period vs.~the final mass of the donor star is shown in Fig.~\ref{fig:fig10}.  The dense locus of steeply rising points maps out the well-known relation between the final orbital period and the mass of the relic white dwarf that was the core of the evolving giant (see, e.g., Rappaport et al.~1995; Tauris \& van den Heuvel 2014).

We have confirmed and updated the $P_{\rm orb}(M_{\rm wd})$ relation for systems containing a white dwarf whose progenitor has stably transferred its envelope to a companion star, in this case another white dwarf.  The expression that best fits our locus of points where the binary evolution tracks terminate along the giant branch is given by:
\begin{equation}
P_{\rm orb}(M_{\rm wd}) \simeq \frac{4.9 \times 10^6 m_{\rm wd}^9}{\left(1+10 m_{\rm wd}^{3} + 10 m_{\rm wd}^6\right)^{3/2}}  ~{\rm days}
\label{eqn:PM}
\end{equation}
where the mass of the white dwarf is expressed in $M_\odot$, and $2 \lesssim P_{\rm orb} \lesssim 40$ days. This is very close to the expression found by Lin et al.~(2011) for accreting neutron-star systems. The main differences between our treatment and the latter study are that Lin et al.~assumed accretor masses $\ge 1.4 \, M_\odot$ (i.e., neutron stars) and they allowed mass transfer to be partially conservative.

The white dwarf relics of the donor stars in Fig.~\ref{fig:fig10} are color coded according to their central chemical composition.  The black color indicates a nearly pure He composition with the C+O fraction less than 5\% by mass. These are largely confined to masses of $\lesssim 0.32 \, M_\odot$.  There are higher-mass He white dwarf relics of the donor star, but they follow the $P_{\rm orb}(M_{\rm wd})$ relation given in Eqn.~(\ref{eqn:PM}) and lie off the top of the plot.  By contrast, the red colored points indicate He/CO hybrid white dwarfs with at least 5\% C+O by mass at the center of the star.  These are descendants of the most massive of the donor stars in our array of initial system parameters, i.e., with $2.2 \lesssim M_{2,i} \lesssim 4 \, M_\odot$.  Such stars do not follow the simple core mass--radius relation of lower-mass stars.

\section{Orbital Period Distribution}
\label{sec:dist}

\begin{figure}
\centering
\includegraphics[angle=0, width=\columnwidth]{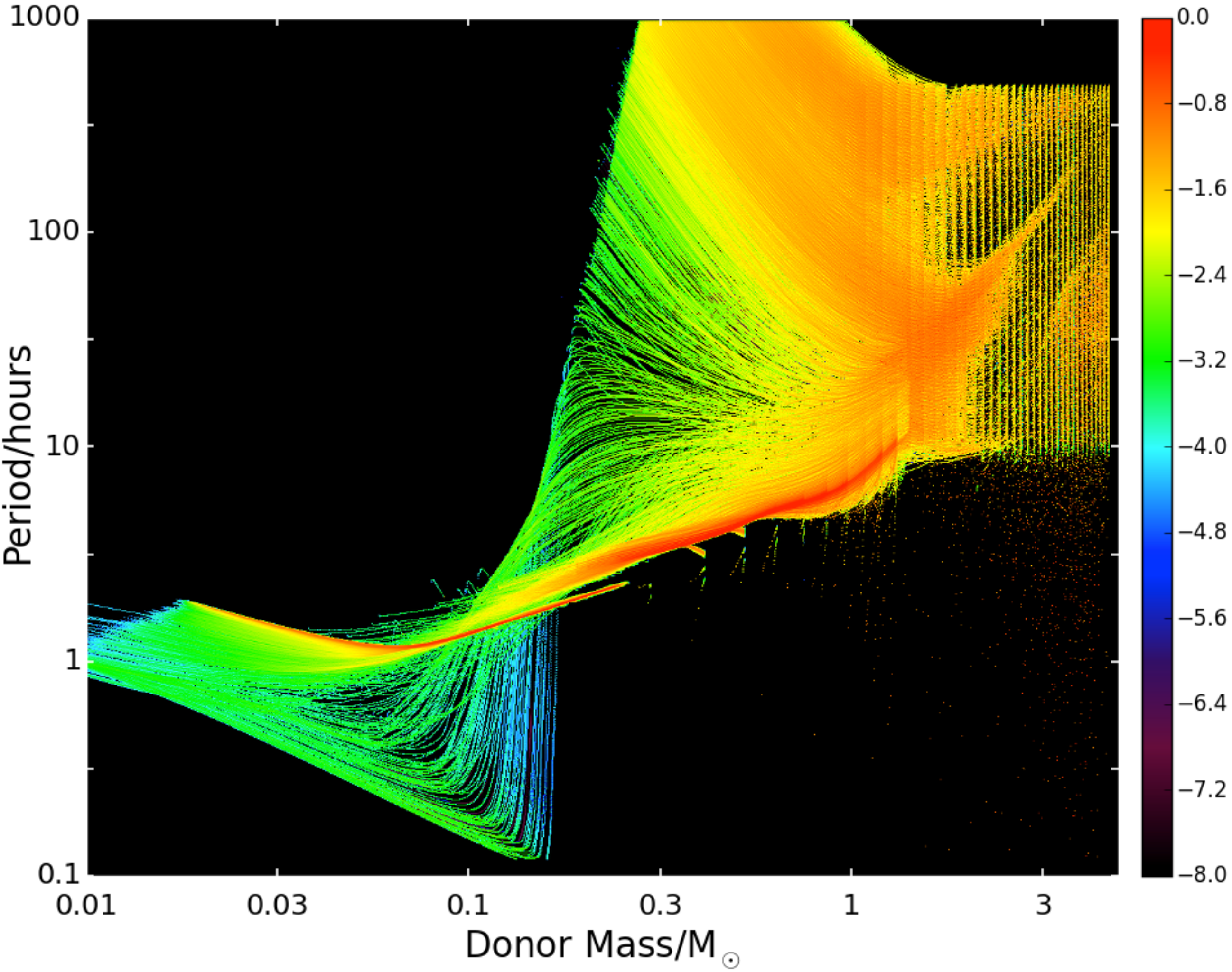}
\caption{Evolution tracks in the $P_{\rm orb}-M_{\rm don}$ plane, color coded by the accumulated mass lost by all the donor stars within a given pixel.  The color scale is logarithmic in mass with red corresponding to $\Delta M_{\rm don} \simeq 1 \, M_\odot$, while green represents $\Delta M_{\rm don} \simeq 10^{-3} \,M_\odot$, and blue implies $\Delta M_{\rm don} \simeq 10^{-5} \,M_\odot$.}
 \label{fig:dM}
\end{figure}

\begin{figure}
    \centering
    \includegraphics[angle=0, width=\columnwidth]{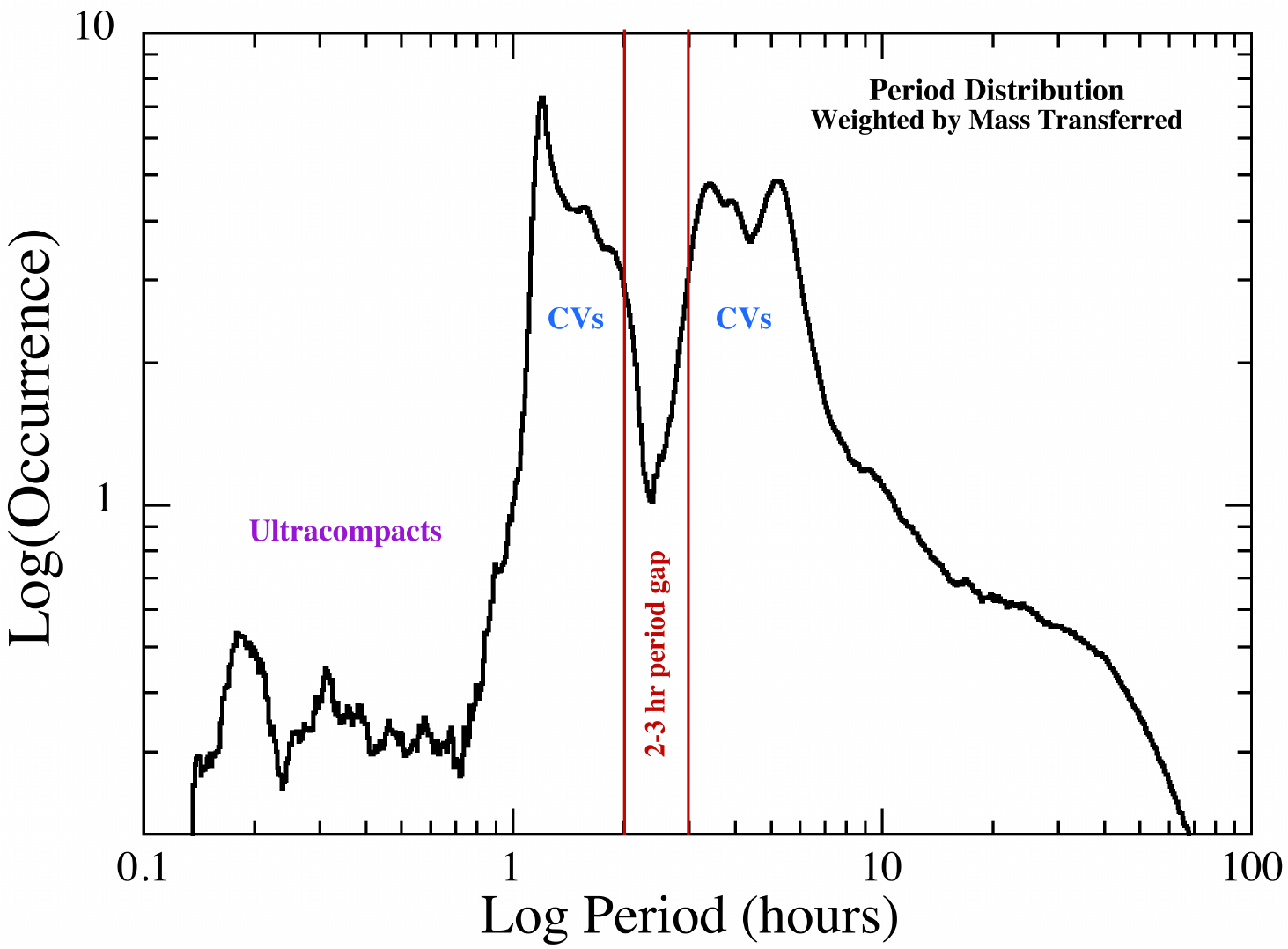}
    \caption{Distribution of orbital periods for all evolution tracks.  The weighting is given by the mass that is lost from the donor star in each evolutionary time step (see Eqn.~\ref{eqn:weight2}). }
  \label{fig:Pdist}
\end{figure}

In order to learn more about the expected orbital period distribution of binary systems with accreting white dwarfs, we can collapse any of the 2D images of the $P_{\rm orb}-M_{\rm don}$ plane onto the $P_{\rm orb}$ axis by integrating over the $M_{\rm don}$ dimension.
The question is - what is the most appropriate weighting function, $W(P_{\rm orb},\dot M_{\rm don})$?  Following the line of arguments given by RJW, we might take
\begin{equation}
W(P_{\rm orb},\dot M_{\rm don}) ~\propto~ \frac{\mathcal{L}_{\rm opt}^{3/2}}{| \dot P_{\rm orb} |}
\label{eqn:weight1}
\end{equation}
where $\mathcal{L}_{\rm opt}$ is the luminosity in the optical band. The 3/2 power simply reflects the increasing volume over which a flux-limited sample can be detected.  The $| \dot P_{\rm orb} |^{-1}$ term reflects the amount of time a system spends per interval of orbital period.  Exactly how the optical luminosity, $\mathcal{L}_{\rm opt}$, depends on $\dot M$ is debatable since some, but not all, of the accreting WD systems are transients.  For simplicity we adopt the optical luminosity from a steady-state Shakura-Sunyaev (1973) disk which scales as $\dot M^{2/3}$ in any fixed passband.  In this case,
\begin{equation}
W(P_{\rm orb},\dot M_{\rm don}) ~\propto~ \frac{|\dot M|}{| \dot P_{\rm orb} |} ~\propto~ \frac{|\Delta M|}{| \Delta P_{\rm orb} |} ~\propto~ \frac{dM/d\ln P_{\rm orb}}{P_{\rm orb} }
\label{eqn:weight2}
\end{equation}
Since in all of our 2D images, the $P_{\rm orb}$ axis is inherently logarithmic in its binning, the weighting can be accomplished simply by summing over the columns along rows in a $P_{\rm orb}-M_{\rm don}$ plot that has been weighted by the amount of mass transferred per evolutionary time step.  Once this collapse onto the $P_{\rm orb}$ axis has been done, we simply divide the sum for each row by the value of $P_{\rm orb}$ for that row.

The evolution tracks in the $P_{\rm orb}-M_{\rm don}$ plane weighted by the mass transferred per evolutionary step are shown in Fig.~\ref{fig:dM}.  After collapsing this image onto the $P_{\rm orb}$ axis, we divide each logarithmically spaced interval of orbital period by $P_{\rm orb}$, and we thereby produce the orbital period distribution shown in Fig.~\ref{fig:Pdist}.

\begin{figure}
    \centering
    \includegraphics[angle=0, width=\columnwidth]{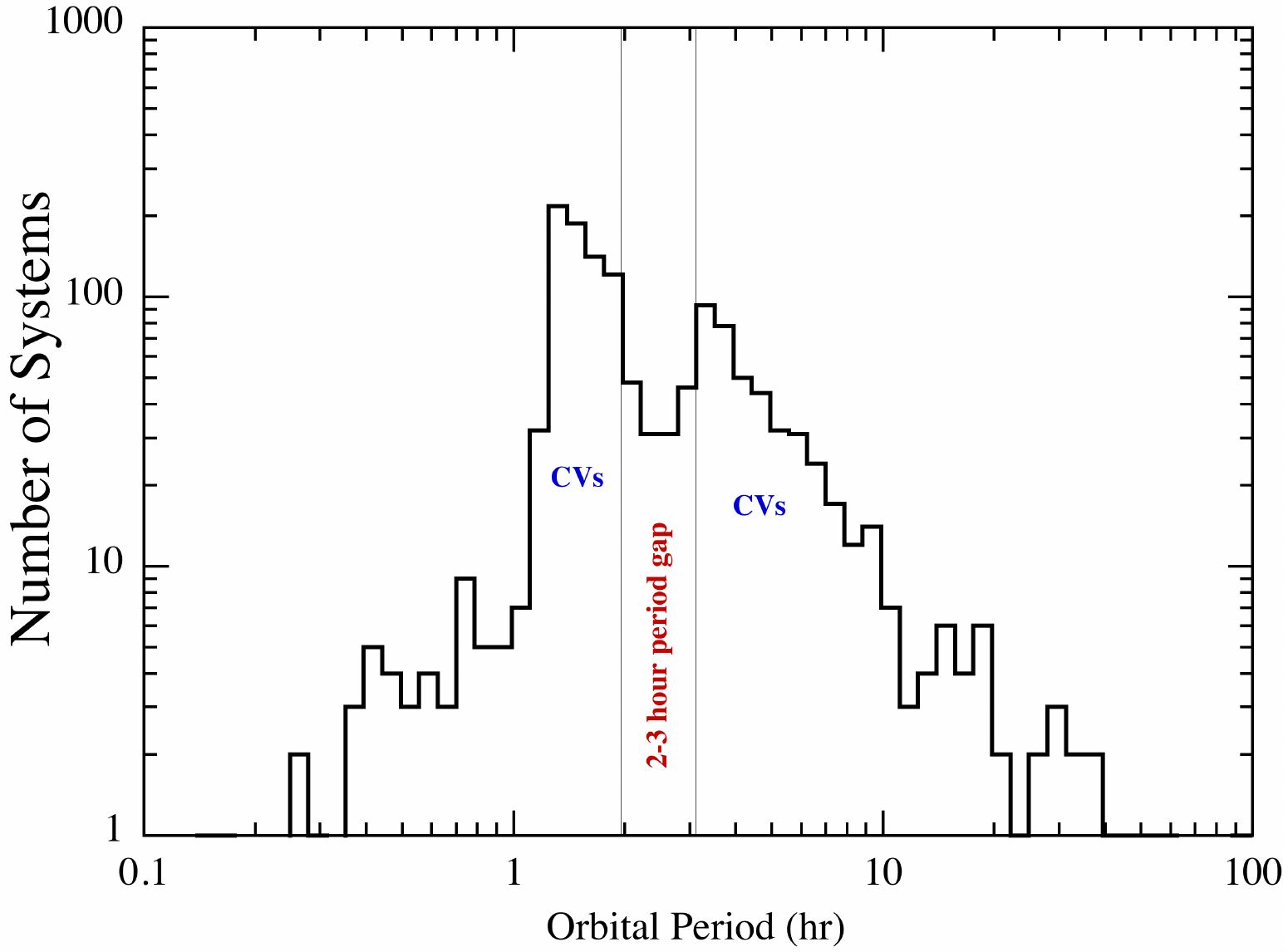}
    \caption{Distribution of orbital periods from the Ritter \& Kolb (2003) catalog.  The notation is similar to that in Fig.~\ref{fig:Pdist}.}
  \label{fig:Ritter}
\end{figure}

This orbital period distribution has several notable attributes and features. First, the distribution falls off dramatically for $P_{\rm orb} \gtrsim 40$ hours and for $P_{\rm orb} \lesssim 50$ minutes (as is observed in Nature; see Sect.~\ref{sec:discuss} and Fig.~\ref{fig:Ritter}).  Second, we notice the prominent `period gap' between $\sim$2 and 3 hours.  The exact range of the theoretical period gap depends somewhat on the specific choice of the parameters used in the magnetic braking law (e.g., Goliasch \& Nelson 2015).  Third, there is no large prominent `spike' at the orbital period minimum of $\sim$70 minutes as discussed by Kolb \& Baraffe (1999) and King et al.~(2002).  The lack of such a spike results from the ranges of chemical compositions and masses of the donor stars as well as the different masses of the accreting white dwarfs, which are all averaged over in the evolutionary ensemble.  Finally, the UC systems that are very apparent in the $P_{\rm orb}-M_{\rm don}$ images, nonetheless have a rather small likelihood in the $P_{\rm orb}$ distribution.   This is due to the fact, discussed in Sect.~\ref{sec:illustrative} (see Eqn.~\ref{eqn:bifur}), that the formation of such systems requires a finely tuned set of initial conditions, in particular an orbital period that is very close to the bifurcation limit at the start of mass transfer.

\section{Accretion Disk Stability}
\label{sec:disk_stability}

\begin{figure}
\centering
\includegraphics[angle=-0, width=\columnwidth]{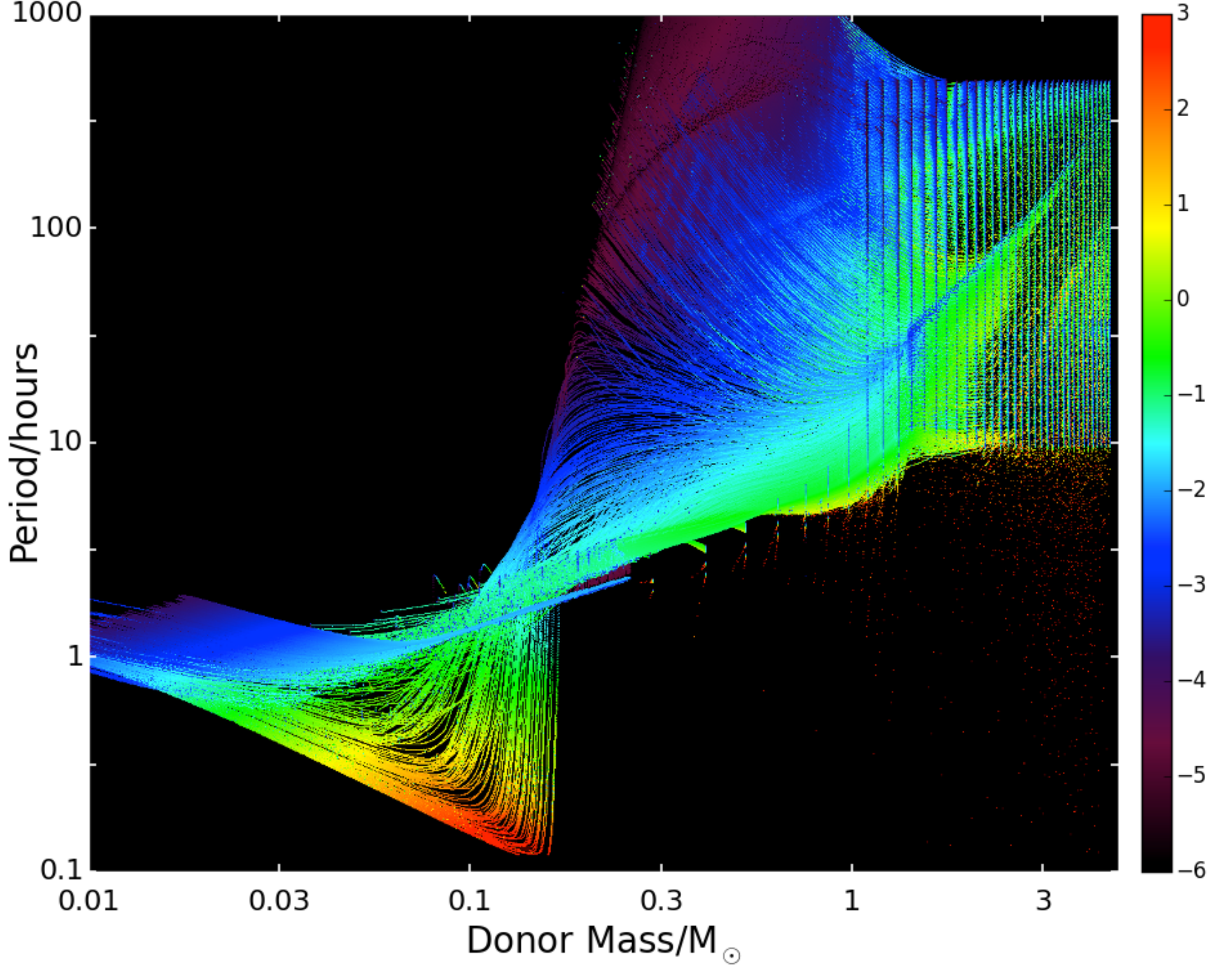}
\caption{Evolution tracks color-coded according to the mean value of $\log \dot{M}/\dot{M}_{\rm crit}$, where $\dot{M}_{\rm crit}$ is the critical mass transfer rate required for stability of the accretion disk against the thermal-viscous instability (see text). Colors are set so that green is near the instability line with ratios of $\dot{M}/\dot{M}_{\rm crit}$ between 0.1-1. Blue and purple regions should have unstable disks; the red regions should be stable. }
 \label{fig:disk}
\end{figure}

The transient nature of most CV-like systems is thought to be the result of the thermal-viscous disk instability mechanism (or `DIM'; e.g., Lin et al.~1985; Cannizzo 1986; Hameury et al.~1998; Lasota 2001).  In CV-like systems, i.e., with white-dwarf accretors, it is currently believed that the dominant effect in determining whether the disk is subject to DIMs is the mass transfer rate, rather than the disk irradiation temperature.  The latter is more likely to be important for LMXBs (for a comprehensive review, see Lasota 2001).   The basic governing equation is (A.4) in Lasota (2001; see also Eqn.~39 of Hameury et al.~1998), which gives the critical value of $\dot M$ at radius $r$ in the accretion disk.  Lasota (2001) then uses this expression to derive his Eqn.~(35) which is the critical $\dot M$ required for overall disk stability, including at the outer reaches of the disk.  After correcting for an incorrect Roche-lobe radius (Eqn.~(34) in Lasota 2001), we find the following closely related expression for the critical $\dot M$ required for stability:
\begin{equation}
\dot M_{\rm crit} \simeq 3.3 \times 10^{-9} f(q)^{2.68} (M_{\rm bin}/M_{\rm wd})^{0.89} P_{\rm orb,hr}^{1.79}   M_\odot ~{\rm yr}^{-1}
\end{equation}
where $f(q)$ is Eggleton's (1983) expression for the scaled size of the Roche lobe, $q \equiv M_{\rm wd}/M_{\rm don}$, and $M_{\rm bin} \equiv M_{\rm wd}+M_{\rm don}$.  We have omitted the dependence on $\alpha$, the usual Shakura \& Sunyaev (1973) viscosity parameter, because it goes only as the 1/100th power.

We compute $\dot M_{\rm crit}$ at every point along all evolutionary tracks. We then take the ratio of the MESA-calculated value of $\dot M$ for the model to $\dot M_{\rm crit}$.  The log of this ratio at all points in the $P_{\rm orb}-M_{\rm don}$ plane is shown in Fig.~\ref{fig:disk}. The colors are set so that green is near the instability line with ratios of $\dot M/\dot M_{\rm crit}$ between 0.1-1.  Normal CV tracks lie in the green region of the diagram.  For all blue and purple regions, including for normal CVs past the minimum orbital period and giant-branch systems, the accretion disks are likely unstable against the thermal-viscous instability.  The only regions of the diagram where the accretion disks are likely to be stable are for ultracompact binaries with $P_{\rm orb} \lesssim 0.3$ hr (18 min).  The SXS region, near $M_{\rm don} \simeq 2 \, M_\odot$ and $P_{\rm orb} \simeq 10-50$ hours, is mostly green in this plot, and it is therefore not obvious whether these accretion disks would be subject to the thermal-viscous instability or not.  However, the very high X-ray luminosities of these sources (i.e., $\gtrsim 10^{37}$ ergs s$^{-1}$) may also help to stabilize the disks via irradiation.

Interestingly, the Palomar Transient Factor has discovered a number of AM CVn binaries by explicitly searching for transient systems (Levitan et al.~2013).  The ones they have discovered all have $P_{\rm orb} \gtrsim 26$ min, in agreement with our prediction that the shortest period AM CVn systems may well be non-transient systems.

\section{Gravitational Radiation}
\label{sec:gravity}

\begin{figure}
\centering
\includegraphics[angle=0, width=\columnwidth]{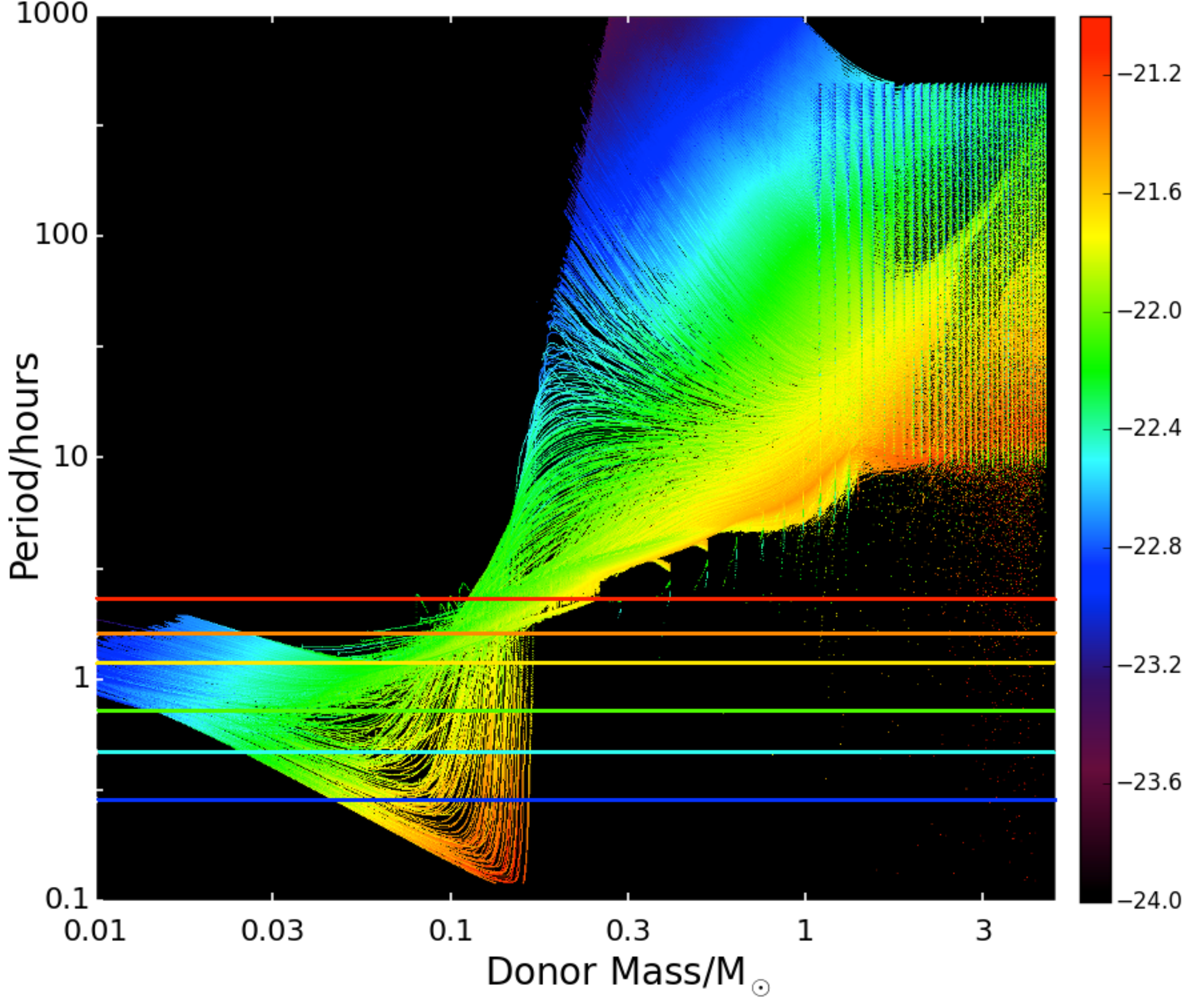}
\caption{The median logarithm of the gravitational wave strain for an assumed distance of 5 kpc for all phases of evolution in the $P_{\rm orb}-M_{\rm don}$ plane.  If we assume that these binaries are observed with LISA at some future date, with integration times of 1 month, the limiting strain sensitivities are given by the horizontal lines.  The color of the line corresponds to the limiting sensitivity at that $P_{\rm orb}$.  The sense of these lines is that any color in the diagram below a horizontal line of the same color is potentially detectable with LISA.}
 \label{fig:GWs}
\end{figure}

As has been noted by a number of authors (Nelemans et al.~2004; Postnow \& Yungelson~2006), some of the shorter period CV-like systems, especially the AM CVn systems, are excellent potential sources of gravitational radiation that would fall within the LISA frequency band of high sensitivities (e.g., Prince et al.~2002; Thrane \& Romano 2013).  We can apply these ideas to the collection of binary evolution tracks in our sample.  The common amplitude for the two GW polarizations, $h_+$ and $h_\times$, is
\begin{equation}
h \simeq \frac{2 (2 \pi)^{2/3}}{D} \frac{G^{5/3}}{c^4} \frac{M_{\rm wd}M_{\rm don}}{(M_{\rm wd}+M_{\rm don})^{1/3}} P_{\rm orb}^{-2/3}
\end{equation}
where $D$ is the distance to the source.  We adopt a fiducial distance of $D= 5$ kpc and then evaluate this expression for each point along all the evolution tracks.  The results are shown in the $P_{\rm orb}-M_{\rm don}$ plane in Fig.~\ref{fig:GWs} with the color coding made according to the logarithm of the gravitational wave strain.

As has been noted before, the best potential sources for gravitational wave detection are the shortest period AM CVn systems.  With the short periods of these systems (down to $\sim$5 minutes), the corresponding gravitational frequency of 0.007 Hz, lies conveniently within the passband of the future LISA mission (e.g., Heinzel \& Danzmann 2014).  In order to assess the plausibility of detection with LISA, we also plot on Fig.~\ref{fig:GWs} five horizontal lines that are color-coded with the anticipated LISA sensitivity at the period where the line is drawn after a 1-month integration time.  Colored regions in the evolution diagram that lie below the horizontal line of the same color would be potentially detectable.  Note that there are no `blue' systems that lie below the blue line, so the oldest ordinary CVs beyond period minimum are not detectable.  There is a small batch of `green' AM CVns that are past minimum period that can be detected, and all of the `red' colored AM CVn systems should be readily detectable.

Population studies have shown there could be about 11000 AM CVn systems in the Galaxy that are potentially detectable with LISA (Nelemans et al.~2004, Roelofs et al.~2007, Ruiter et al.~2010).

\begin{table}
\scriptsize \caption{Additional systems with white dwarf accretors } \label{tab:Observ_sys}
\resizebox{\columnwidth}{!}{
\begin{tabular}{lllll}
\hline
\hline
System	&	Type	&	P (hr)	&	M${_d}$ ($\textrm{M}_{\odot}$)	&	Ref.	  \\
\hline
CP Eri	&	AM CVn	&	0.47	&	0.035	&	1	 \\
Gaia14aae	&	AM CVn	&	0.83	&	0.015	&	2	 \\
 SDSS J174140.49+652638.7	&	AM CVn	&	1.47	&	0.17	&		3 \\
SDSS J075141.18-014120.9	&	AM CVn	&	1.92	&	0.19	&	3	 \\
SDSS J1152+024	&	AM CVn	&	2.40	&	0.44	&	4	 \\
1E 0035.4-7230	&	SXS	&	4.1	&	~0.4	&	 5	 \\
RX J0527.8-6954	&	SXS	&	9.42	&	~1.3	&	6	 \\
CAL 87	&	SXS	&	10.6	&	~0.34	&	7	 \\
TYC6760-497	&	PRE-SXS	&	11.97	&	~1.23	&	8	 \\
RX J0019.8+2156	&	SXS	&	15.85	&	~1.5	&	9 \\
CAL83	&	SXS	&	25.14	&	0.61	&	10	 \\
\hline
\end{tabular}
}
\scriptsize{References for Table~\ref{tab:Observ_sys}.~1-Armstrong et al.~(2012), 2-Campbell et al.~(2015), 3-Kilic et al.~(2014), 4-Hallakoun et al.~(2016), 5-Kahabka et al.~(1999), 6-Alcock et al.~(1997),
7-Ablimit \& Li~(2015), 8-Parsons et al.~(2015), 9-Becker et al.~(1998), 10-Odendaal et al.~(2015).}
\label{tbl:observ}
\end{table}

\begin{figure}
\centering
\includegraphics[angle=0, width=\columnwidth]{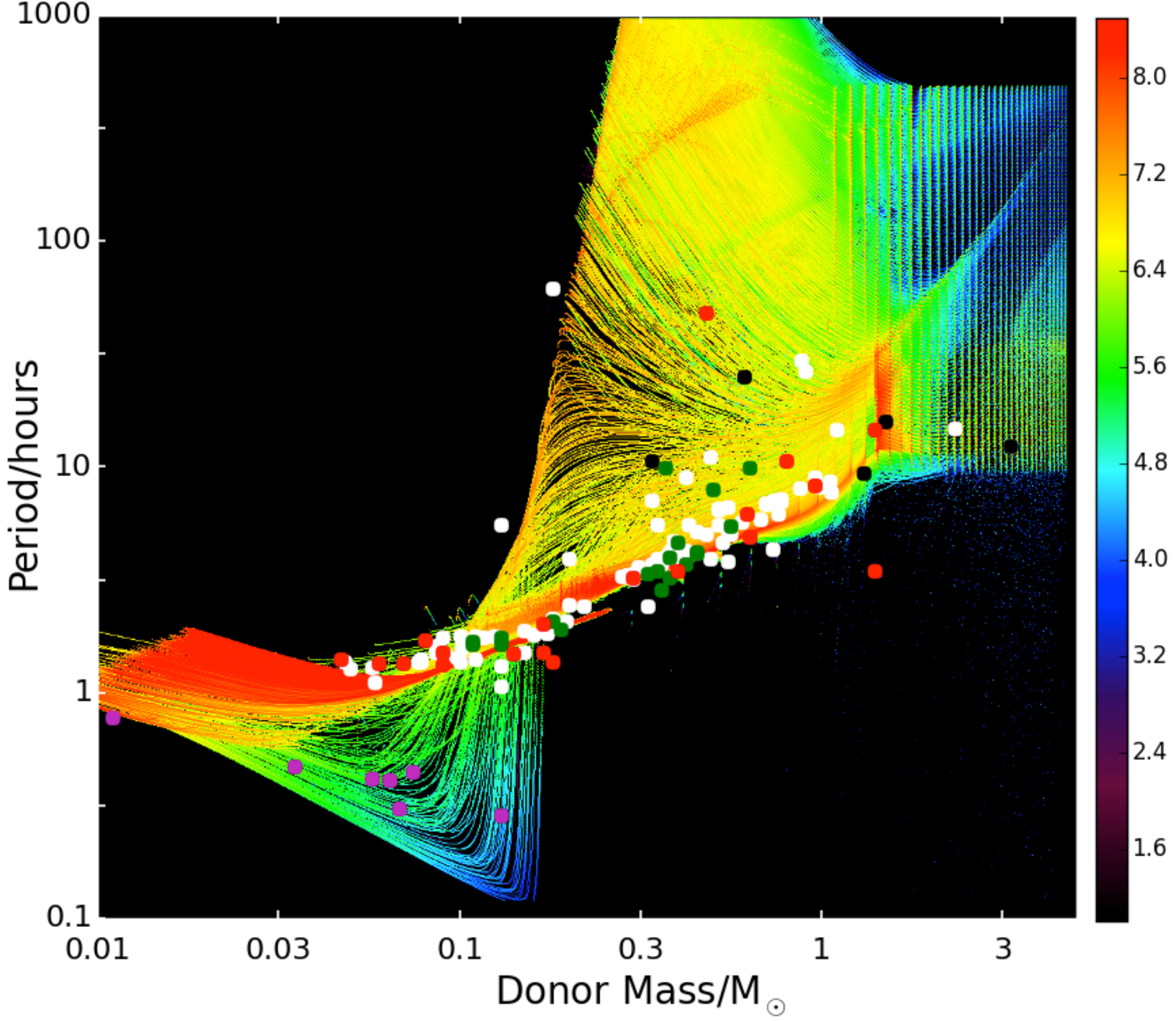}
\caption{Observed systems with accreting white dwarfs where both the orbital period and the donor-star mass are known. Filled white (red) circles stand for CVs with more (less) certain parameters; green for magnetic CVs; magenta for AM CVn; and black for SXS. The systems are superposed on the evolution-time weighted tracks in the $P_{\rm orb}-M_{\rm don}$ plane (essentially the same as shown in Fig.~\ref{fig:time}). An order of magnitude more systems have measured orbital periods covering a wide range, but the donor masses have not been determined (Ritter \& Kolb 2003).}
 \label{fig:observ}
\end{figure}

\begin{figure}
\centering
\includegraphics[angle=-0, width=\columnwidth]{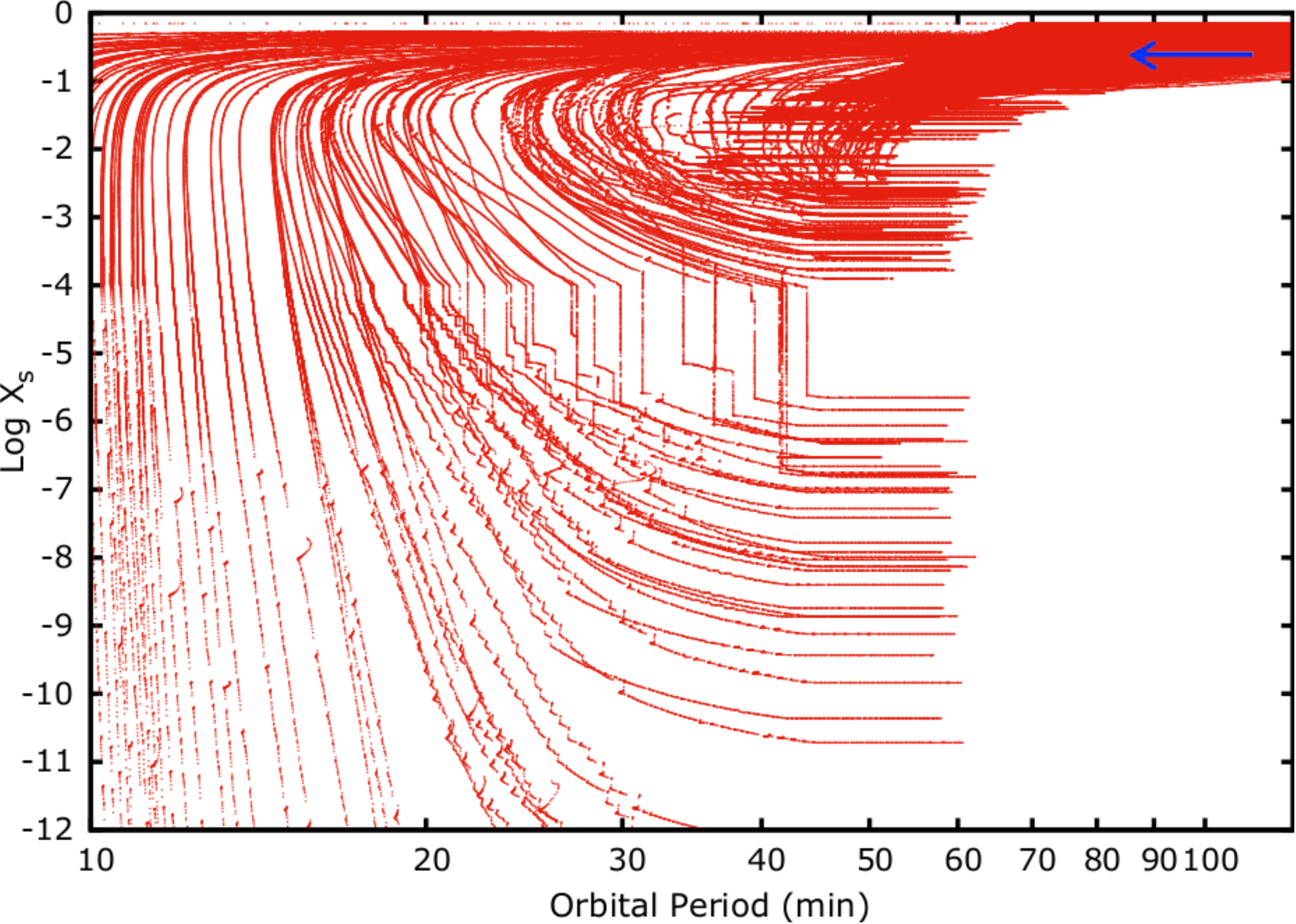}
\caption{Evolution tracks in the plane of surface abundance (by mass) of hydrogen, $X_s$, vs.~the orbital period. The blue arrow indicates the temporal direction of evolution of systems into this part of the diagram from longer periods.  Each evolution track is shown as a set of discrete points, one for each evolution step.}
 \label{fig:Hs}
\end{figure}

\section{Comparison with Observed Systems}
\label{sec:discuss}

There are a very large number of known binaries containing accreting white dwarfs.  Many of these are cataloged by Ritter \& Kolb (2003).  Most of these have known orbital periods (see the orbital period distribution in Fig.~\ref{fig:Ritter}), but the number of them with reasonably well-determined donor masses is much smaller.  Therefore, they are not easily placed in the $P_{\rm orb}-M_{\rm don}$ plane that we have explored in this work.  The CVs for which both $P_{\rm orb}$ and $M_{\rm don}$ are reasonably-well known are plotted as filled white circles on Fig.~\ref{fig:observ} (evolutionary dwell-time weighted diagram).  The AM CVn systems are plotted as filled purple circles.  In general, the SXSs are incompletely listed in the Ritter \& Kolb (2003) catalog, and we have listed 5 additional SXSs in Table \ref{tbl:observ}, with references to their appearance in the literature.  The SXSs are plotted as filled black circles.  Table \ref{tbl:observ} also lists 5 AM CVn systems not found in the Riter \& Kolb (2003) catalog.

In general, there is good overlap between the evolution tracks in Fig.~\ref{fig:observ} and the locations of the circles representing the various types of systems with accreting white dwarfs.  The CVs nicely follow the CV-like evolution tracks in the $P_{\rm orb}-M_{\rm don}$ diagram.  However, there is a distinct dearth of them near the post-minimum period systems with $M_{\rm don} \simeq 0.01-0.03 \,M_\odot$, where in fact, the model CV tracks spend most of their lifetimes.  This is most likely due to a combination of observational bias against detecting systems with very low $\dot M$ and possibly very long outburst cycle times (see, e.g., Ciardi et al.~2006), and the difficulty in directly determining the mass of such low-mass donors.

The SXSs, though few in number, are indeed found in just about the locations in the $P_{\rm orb}-M_{\rm don}$ plane that stand out in Fig.~\ref{fig:SXS}.

The AM CVn systems represent one of the more interesting and important endpoints of binary star evolution. With their short orbital periods (down to $\lesssim 6$ min.), these systems are of considerable interest in gravitational wave astronomy of the future with LISA (Marsh \& Steeghs~2002; Nelemans et al.~2004).  Their high accretion rates of  $10^{-6} -10^{-8}$ $M_{\odot}$ yr$^{-1}$ make them also detectable as X-ray sources (Nelemans et al.~2004).

There have been three formation scenarios proposed for AM CVn systems. (i) Systems consisting of a double white dwarf in close orbits, perhaps the end product of two common envelope phases,  emit gravitational radiation and evolve to very short orbital periods.  At orbital periods of $\sim$2-3 minutes, the stars can commence mass transfer from the less massive to the more massive WD, and if the mass transfer is stable, the systems will evolve to longer orbital periods and possibly resemble an AM CVn (see, Nelemans et al.~2001b).  (ii) In a second scenario a low-mass helium-burning donor, the remnant core of an evolved star that had lost its envelope (perhaps in a CE phase), starts to transfer mass to a white dwarf companion. When the donor star becomes semi-degenerate, the orbital period reaches a minimum, after which the orbit starts to expand (Iben \& Tutukov~1987; 1991). (iii) The third proposed scenario for forming AM CVn systems is from CVs with evolved donor stars (Nelson \& Rappaport~2003; Podsiadlowski, Han \& Rappaport~2003; Nelemans et al.~2010, and references therein).

In the first two of the above scenarios for forming AM CVn systems, the advanced evolutionary stages of stellar evolution that lead to either a HeWD or hybrid COWD (in the first scenario) or a He burning star (in the second scenario), imply very high temperatures near their centers (i.e., $\gtrsim 10^8$ K).  In turn such high temperatures could, for some cases, lead to something of a reversal of the CNO abundances relative to what we see in our late evolutionary phases (i.e., C,O rich and N deficient).\footnote {It is also possible for the CNO abundances in He-burning stars to be close to their nuclear quasi-equilibrium isotopic values.} In particular, Nelemans et al.~(2010) have shown that the ratio of C/N in conjunction with C/O is a good discriminator between scenarios (i) and (ii), described above. Unfortunately it is not simple to use the abundance ratios to discriminate between WD donors, i.e., in scenario (i), and our donors that have evolved from high-mass stars according to scenario (iii). However, if any H is detected (no matter how small) then scenario (iii) would correctly describe the progenitor evolution (see Fig.~\ref{fig:Hs}). Moreover, if N/O can be determined with some confidence to within a factor of two, then this measurement can act as a discriminant between scenarios (i) and (iii).  We would expect the N/O ratio to be at least a factor of two higher for HeWD donors (and it could be as large as an order of magnitude higher).  Given sufficient accuracy in the inferences for each of the H, C, N, and O abundances, it should be possible to break the degeneracy and determine which scenario is operative.

A number of groups have attempted to measure the chemical composition in CV donor stars via direct spectroscopic observations in the K-band region (Harrison et al.~2004; Harrison et al.~2005;  Howell et al.~2010; Hamilton et al.~2011). They collected spectroscopic information on 19 pre-CVs, 31 non-magnetic CVs, and 11 magnetic CVs.  Eighteen of the nineteen pre-CV systems observed show CO commensurate with their spectral type, indicating that the measurements are likely valid and reliable. Among the 19 non-magnetic CVs above the period gap, 10 show either no CO band structures or weakened CO features (relative to what would be expected for the given spectral type), while 4 systems appear largely normal. For CVs below the period gap 3 of 7 clearly show CO bands that are weaker than expected for their spectral type. These results appear to be somewhat counter to the sense of what one might infer based on the results in Figs.~\ref{fig:CO}, \ref{fig:NO}, \ref{fig:CN}.  However, it is important to note in regard to the surface chemical abundances, that only one of these systems below the period gap has an orbital period that is marginally less than the minimum period that would be expected for an unevolved (i.e., H-rich) CV donor.  More recently, Harrison \& Hamilton (2015) investigated three CVs and found significant deficits of carbon as would be expected for donors that descended from high-mass stars. Most recently, Kennedy et al.~(2015) found that CSS1111 showed an anomalously strong N/C ratio that might be indicative of a CV that is evolving to become an AM CVn.  Our impression of these abundance measurements is that the statistics are still too limited and the selection effects too large to draw any robust conclusions.

Not only have we created a mapping for abundance ratios that can be used to understand the possible progenitor evolution and subsequent fate of CV systems, but we also identified a diagnostic to determine whether a `normal' CV in the post-period gap phase of its evolution might have actually descended from a higher mass star.  As can be seen in Figs.~\ref{fig:CO}, \ref{fig:NO}, \ref{fig:CN}, there is a thin uppermost branch of the post-gap systems which have nearly primordial surface compositions (these are all CVs that descended from low-mass donors of $\lesssim 0.8 M_\odot$). The real inferential test of these calculations will occur once the chemical composition of the systems lying just below this main branch of CVs is determined. For a given donor mass, the CVs that evolved from more massive donors will have orbital periods only a few minutes less than their more `standard' cousins but will have C/O and C/N ratios that will be nearly two orders of magnitude smaller. Such a huge difference should, in principle, be readily measured and complications due to factors such as metallicity will not play a role in affecting the conclusions as to the nature of the progenitors.

\section{Summary and Conclusions}
\label{sec:summary}

In Sects.~\ref{sec:results}, \ref{sec:Pmin}, \ref{sec:final}, and \ref{sec:dist} we described the results found from following the binary evolution of 56,000 systems with accreting white dwarfs.  By systematically covering a dense grid of initial values of $M_{\rm wd}$, $M_{\rm don}$, and $P_{\rm orb}$ we were able to explore the evolution tracks of a wide range of possible binary systems with white dwarf accretors -- some already well-studied, but perhaps highlighting a number of regions in the $P_{\rm orb}-M_{\rm don}$ plane where little detailed study has been done.

\noindent
$\bullet$ Our work predicts CV-like systems with periods of 2--40 days and donor masses of $0.3-0.8 \, M_\odot$.  There are only four systems listed in the Ritter \& Kolb catalog with $P_{\rm orb} \gtrsim 2$ day, and only one of those has $M_{\rm don}$ determined.  We urge renewed effort to determine the masses in the other three of those systems, and for observers to be cognizant of the possibility of longer-period CVs. \\
$\bullet$ Only one system is currently known that nearly corresponds to the `transition' systems in our study with $P_{\rm orb} \simeq 5-50$ hours and $M_{\rm don} \simeq 0.2 - 0.4 \,M_\odot$ (see Fig.~\ref{fig:observ}).  As can been seen from Fig.~\ref{fig:observ}, the evolution tracks in this region are sparse, but perhaps sufficient to yield some observable systems.  \\
$\bullet$ The densely populated region with $P_{\rm orb} \simeq 1-2$ hr and $M_{\rm don} \simeq 0.01-0.03 \,M_\odot$ may contain many post period minimum CVs with very small $\dot M$.  However, there is a serious need for measurements of the donor star mass in short-period CVs, and possibly their chemical composition in order to verify the existence of this type of object.

We have explored the range of CV tracks with different evolutionary states of the donor star.  We find that the minimum orbital period varies continuously from 70 minutes down to below 40 minutes, depending on the H fraction in the core of the donor (see Sect.~\ref{sec:Pmin}). This spread in values for $P_{\rm min, orb}$ due to the different chemical compositions for the donor star, as well as the range of masses for the WD accretors, tends to remove any spikes in the distribution of $P_{\rm orb}$ at orbital period minimum (see Sect.~\ref{sec:dist}).

We have also evaluated the range where SXS sources are likely to be found, namely with $P_{\rm orb}$ in the range of 4-50 hours, and donor masses in the range of 0.5-1.5 $M_\odot$.  CAL 83, CAL 87, RX J0019.8+2156, and RX J0925.7-4758 have orbital periods of 25 hr, 11 hr, 16 hr, and $\sim$100 hr, respectively (see \url{http://www.mpe.mpg.de/jcg/sss/ssscat.html}; Motch et al.~1994; Becker et al.~1998; Odendaal et al.~2015; Ablimit \& Li~2015).  To the extent that the donor masses in these systems are known, they are consistent with the theoretically determined range listed here.  We  can, in addition, expect to find SXS with longer periods in the range of 50-200 hours, but with a population that is $\sim$5 times lower than for their shorter-period counterparts.  Finally, several symbiotic novae, including AG Dra, which has a likely orbital period of $\sim$550 days, are also considered to be SXSs.  However, our evolution calculations do not explore to such long orbital periods, nor would mass transfer via a strong stellar wind from the donor star be included in the calculations.

We have also studied the ultracompact binary regime in more detail than has previously been done (Podsiadlowski et al.~2002; Podsiadlowski et al.~2003).  We have reaffirmed that UC systems can indeed evolve from a normal donor star starting in a $\sim$20-30 hour orbit with a WD accretor when mass transfer commences.  However, it is also clear that only a small fraction of white dwarf accreting systems will evolve through this channel because of the finely tuned initial period that is required -- typically only about 1 hour wide for any given set of initial masses.  The relative dearth of these systems can be seen quantitatively in the orbital period distribution in Fig.~\ref{fig:Pdist}.  We note that the MESA code is particularly robust when computing low-mass models of the donor stars during the UC phase because its equation of state realistically incorporates pressure ionization and Coulombic interactions (amongst other interactions).

Finally, we have explored the surface composition of the donor star during all the evolutionary phases.  We make definite predictions about the N/O and N/C ratios in different parts of the $P_{\rm orb}-M_{\rm don}$ plane.  In particular, the very large over-enhancement of N with respect to C and modestly less in O is a potentially important signature for understanding the origin of AM CVn systems and all CVs with $P_{\rm orb} \lesssim 70$ minutes.    A modest enhancement in N/C (by only a factor of 2) in classical CVs, below the period gap, would clearly indicate an origin in more massive and evolved donor stars.  In particular, CVs with $P_{\rm orb} \approx 60 - 70$ min should have N/O enhanced to 0.3, while for $P_{\rm orb} \lesssim 50$ min, the N/O ratio can be as high as unity (these are to be compared with the primordial N/O ratio of 0.1).  Pioneering measurements of the chemical composition of the donor stars in CVs have begun (see, e.g., Harrison \& Hamilton 2015, and references therein; Kennedy et al.~2015), but the sample is still small, and there is only one system with a period just below 70 minutes (at 64 min).  Nonetheless, given the nearly exponential rate of discovery of CVs over the past two decades (see Ritter \& Kolb 2003) we can expect this situation to change.  Even measurements of the C/O and C/N abundance ratios of the donors of systems that have clustered near the minimum orbital period of `normal' CVs will help us answer the question as to what fraction of these donors descended from high-mass progenitors and thus allow us to infer whether the assumptions that are being used in population syntheses are reasonable.

\vspace{0.3cm}
 {\em Note Added in Manuscript:} After this work was largely complete, we ran an additional smaller grid of 1848 models with lower-mass donor stars (including 0.1 and 0.2 $M_\odot$).  We found that those did not change any of our conclusions, and therefore we have not included them in the various plots.

\acknowledgments

We thank Steve Howell for helpful comments about the surface composition of the donor stars in CVs, and Jonas Goliasch for technical discussions.  We are grateful to an anonymous referee for his/her insightful comments and suggestions which significantly helped to improve the manuscript. Computations were made on the supercomputers Mammouth parall\`{e}le II and Mammouth s\'{e}rie II from the Universit\'{e} de Sherbrooke, managed by Calcul Qu\'{e}bec and Compute Canada. The operation of these supercomputers is funded by the Canada Foundation for Innovation (CFI), NanoQu\'{e}bec, RMGA, and the Fonds de recherche du Qu\'{e}bec - Nature et technologies (FRQNT).   B.\,K.  is  grateful  to  the  MIT  Kavli  Institute for  Astrophysics  and  Space  Research and MIT Department of Physics for the hospitality they extended during her visit and gratefully acknowledge the support provided by the Turkish Scientific and Technical Research Council (T\"UB\.ITAK-112T766 and T\"UB\.ITAK-B\.IDEP 2219).  L.\,N. would like to thank the Natural Sciences and Engineering Research Council (NSERC) of Canada for financial support.  K.\,Y.  acknowledge support from the the Turkish Scientific and Technical Research Council (T\"UB\.ITAK-113F097). J\,Q. is supported by the Walter C.\,Sumner Memorial Fellowship and by the Vanier Canada Graduate Scholarship administered by the Natural Sciences and Engineering Research Council of Canada (NSERC).

{MESA (v3851; Paxton et al. 2011; 2013; 2015)}

\appendix
\section{Evolution Tracks in the HR Diagram}
\label{app:drift}

\begin{figure}[h!]
\begin{center}
\includegraphics[width=0.36 \textwidth, height=6cm]{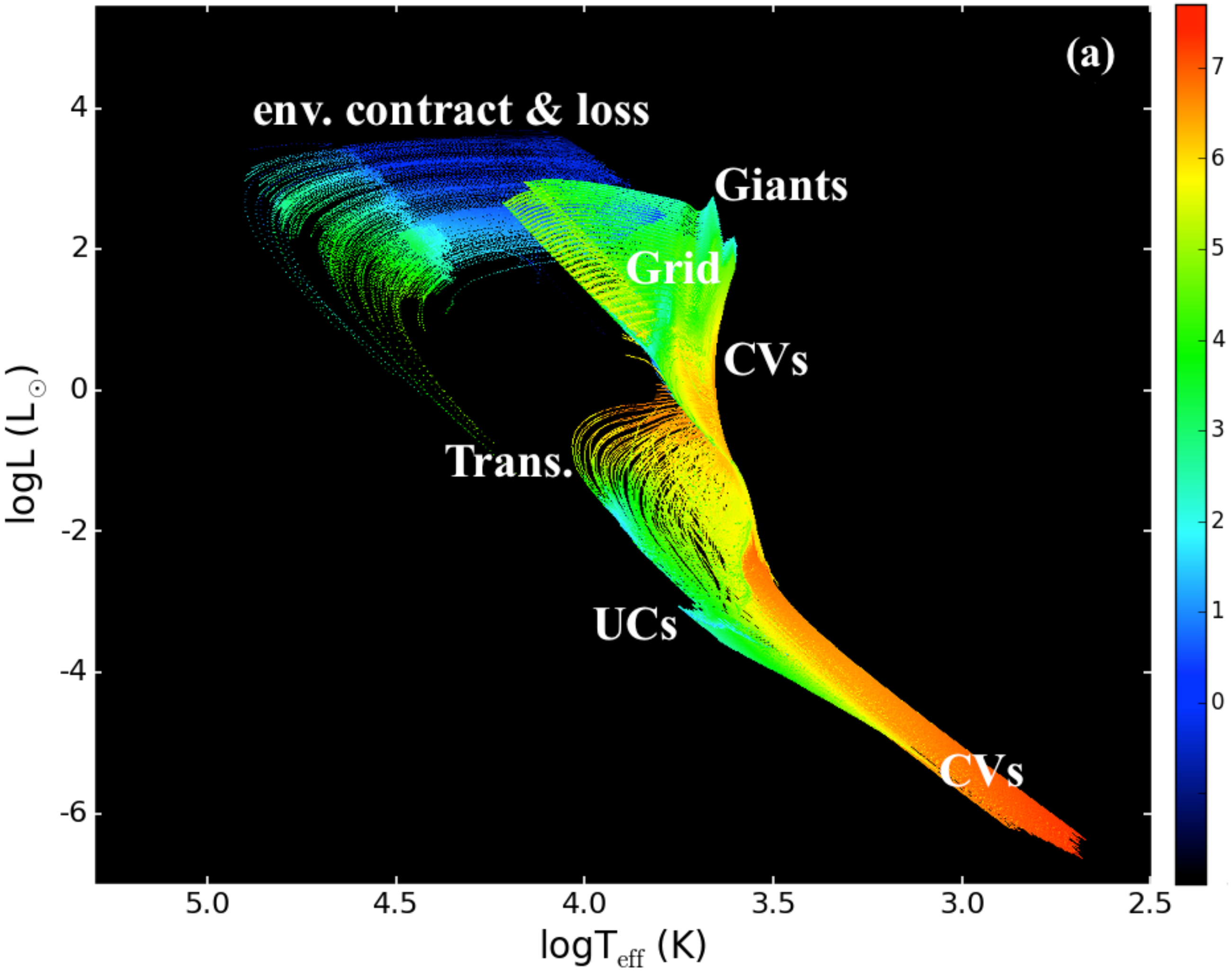} \hglue1.2cm
\includegraphics[width=0.37 \textwidth, height=6.01cm]{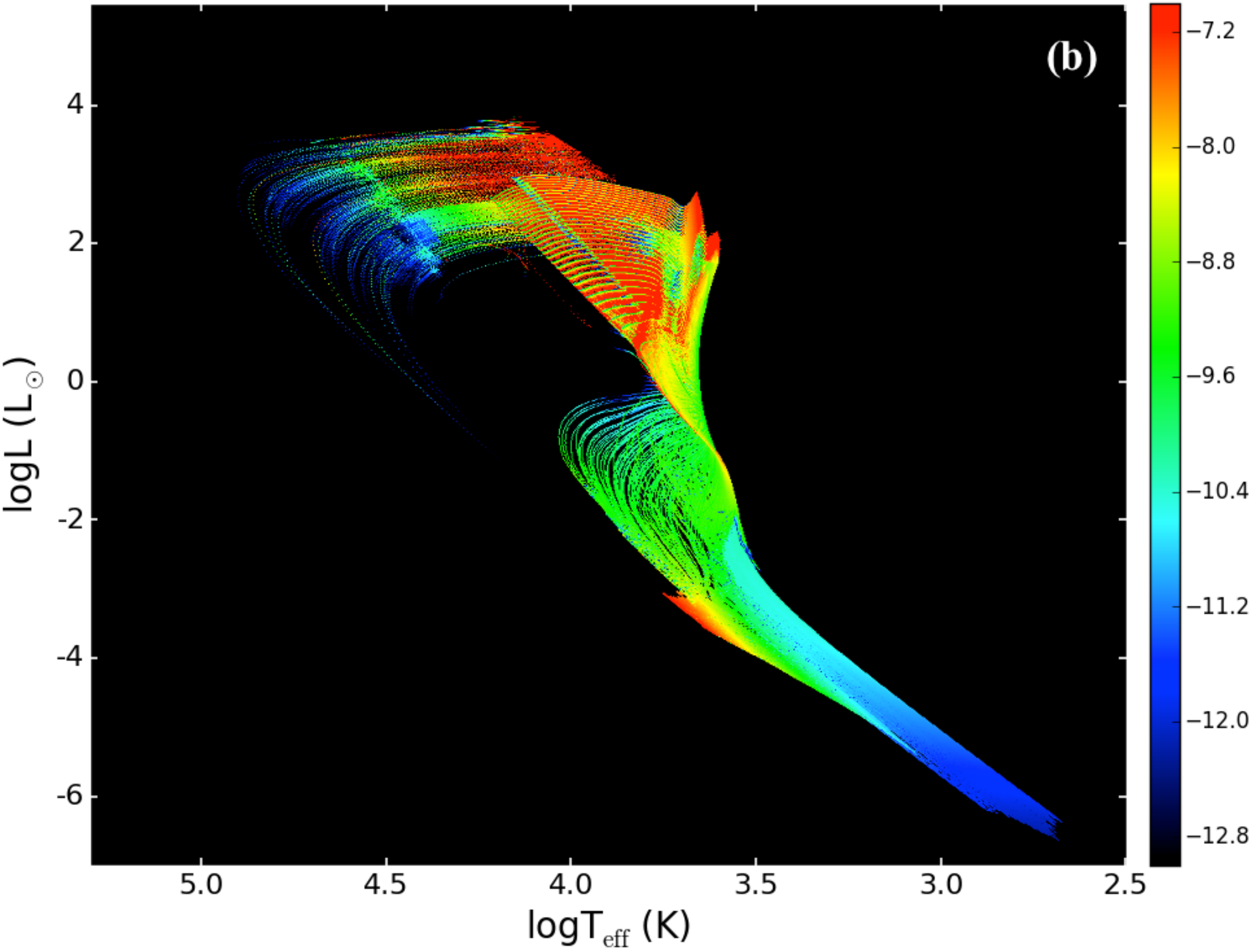} \vglue0.6cm
\includegraphics[width=0.376 \textwidth, height=6.09cm]{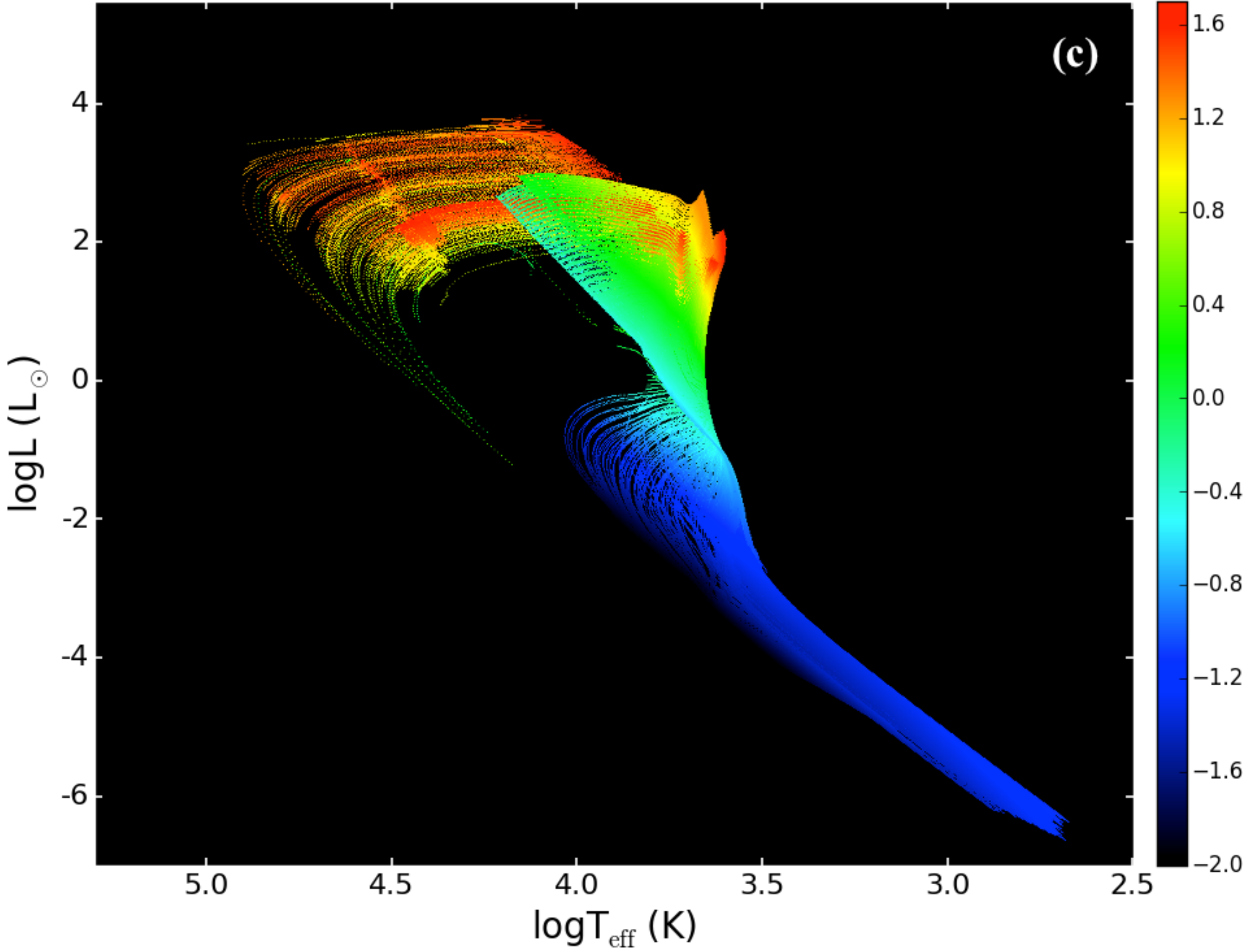} \hglue0.48cm
\includegraphics[width=0.405 \textwidth, height=6.14cm]{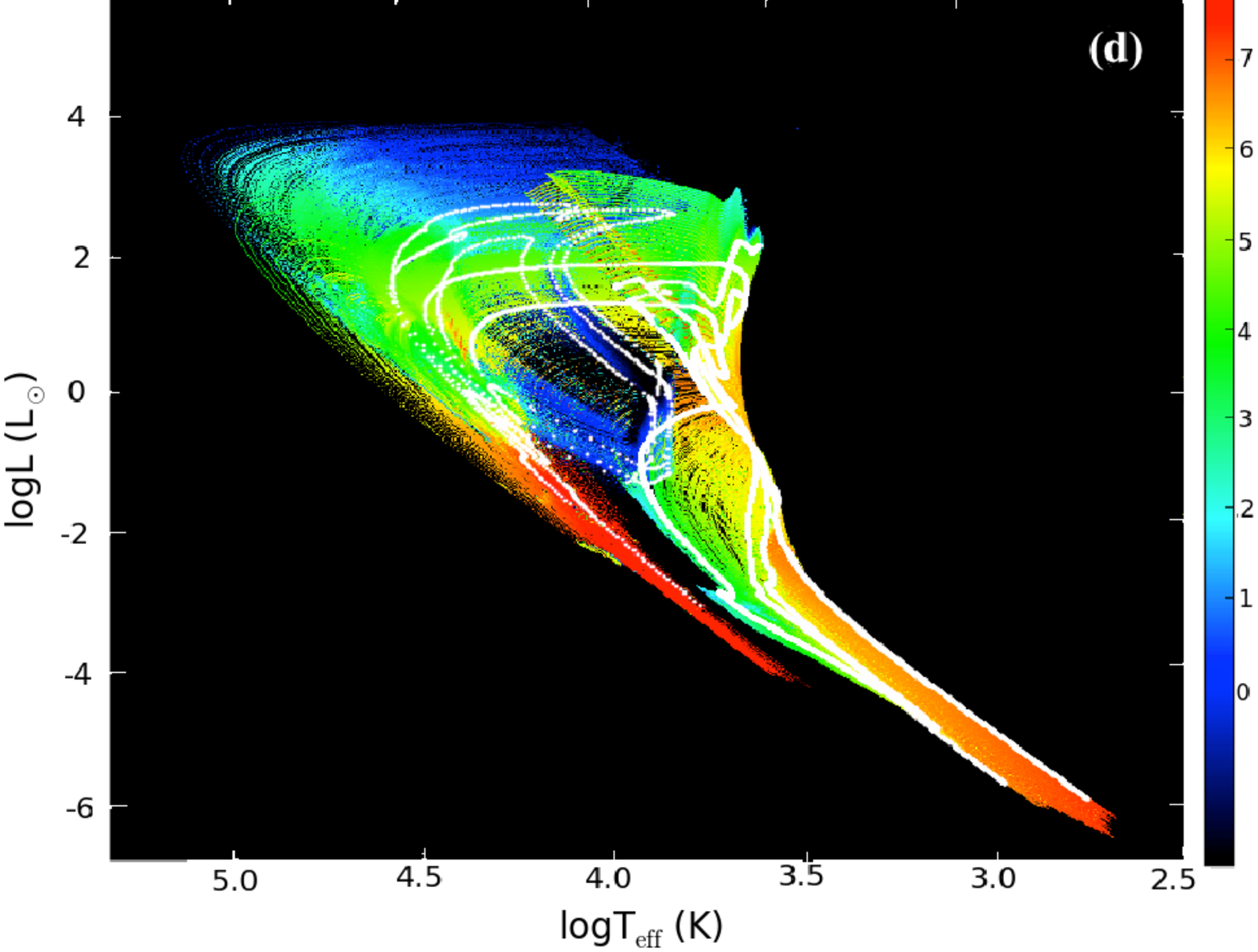} \vglue0.6cm
\caption{Evolution tracks in the Hertzsprung-Russell plane.  Panels (a)-(c) are the tracks during times when $\dot M > 10^{-12} \, M_\odot$ yr$^{-1}$, and with color coding according to the logarithm of the mean dwell time per pixel (panel a), $\dot M$ (panel b), and $P_{\rm orb}$ (panel c).  The bottom right panel (d) contains all the tracks, regardless of $\dot M$, and is color-coded according to the median evolutionary dwell time. The labels in panel (a) refer to cataclysmic variables (`CVs'), ultracompact binaries (`UCs'), initial grid of models (`Grid'), giant donor stars (`Giants'), transition systems (`Trans.') heading toward UCs, and the contraction and loss of the giants' envelope (`env.~contract.~and loss').}
\label{fig:HR}
\end{center}
\end{figure}

In this Appendix we show the evolutionary tracks that were generated for this paper as they would appear in the Hertzsprung-Russell plane of $\log L$ vs.~$T_{\rm eff}$.  The first 3 panels in Fig.~\ref{fig:HR} show the CV tracks only for times when the mass transfer rate exceeds $10^{-14} \,M_\odot$ yr$^{-1}$ (i.e., when the systems might be detectable as mass-transferring binaries).  Those three panels are color-coded by the median evolutionary dwell time, by $\dot M$, and by $P_{\rm orb}$, respectively.  In the fourth panel the evolution tracks are shown for all systems, with or without mass transfer.

It is important to note that these images display the tracks of the donor stars and do not include the light from either the white dwarf accretor or from any possible accretion disk.

In the first panel, we have labeled a few of the key regions of the evolutionary diagram as they correspond to the tracks in the $P_{\rm orb}-M_{\rm don}$ plane  as presented in the rest of the paper.  By necessity, because of the highly non-linear mapping between the two planes, these regional markers are only approximately placed.

Most of the systems are born near the middle of the diagram and the CVs evolve down and to the right, while the giants evolve up and first to the right, and then toward the left as their envelopes become completely lost.  The trend in the periods can clearly be discerned from panel (c), where the color coding is indicative of $P_{\rm orb}$.  The coding according to dwell time (panel a) shows the CV `graveyard' where systems with very low-mass donor stars spend long intervals of time, in contrast with the donors on the giant branch which evolve very rapidly, especially during the time when they have lost most of their envelopes.  In panel (b) which is color coded according to $\dot M$, the CVs with short orbital periods have low mass transfer rates, while stars on the giant branch have high $\dot M$.  Note, the small notch near $\log T_{\rm eff}(K) \simeq 3.7$ and $\log (L/L_\odot) \simeq -3.5$ where the ultra-compact systems are located.

With regard to the ``all-tracks'' plot (panel d), the main new features, not seen on the other panels, are the end points of the evolution tracks of the giant donor stars. The tracks at high, and nearly constant luminosity, moving toward higher $T_{\rm eff}$ are for giants whose envelopes are running out of mass, and are shrinking while getting hotter (for a comprehensive review see Iben 1995).  After they reach a maximum $T_{\rm eff}$, the underlying core begins to cool, ultimately at near constant radius on the white-dwarf cooling branch.

\clearpage

\clearpage

\end{document}